\documentstyle[12pt]{article}
\newfont{\bg}{cmr10 scaled\magstep4}

\newcommand{\bigzerou}{%
   \smash{\lower1.7ex\hbox{\bg 0}}}
\renewcommand{\theequation}{\thesection.\arabic{equation}}

\def\a{\begin{eqnarray}}
\def\b{\end{eqnarray}}
\def\0{\nonumber}
\def\ba{\begin{array}}
\def\ea{\end{array}}

\newlength{\extraspace}
\newlength{\extraspaces}
\newcounter{dummy}

\newcommand{\ai}{
\addtocounter{equation}{1}
\setcounter{dummy}{\value{equation}}
\setcounter{equation}{0}
\renewcommand{\theequation}{\thesection.\arabic{dummy}\alph{equation}}
\begin{eqnarray}
\addtolength{\abovedisplayskip}{\extraspaces}
\addtolength{\belowdisplayskip}{\extraspaces}
\addtolength{\abovedisplayshortskip}{\extraspace}
\addtolength{\belowdisplayshortskip}{\extraspace}}
\newcommand{\bj}{
\end{eqnarray}
\setcounter{equation}{\value{dummy}}
\renewcommand{\theequation}{\thesection.\arabic{equation}}}
\def\noal{\noalign{\vskip10pt}}

\begin{document}

\begin{flushright}
UT-HEP-696-94 \\
hepth/9412226
\end{flushright}

\vskip 1.0cm
\centerline{\Large\bf Calculation of Gromov-Witten  invariants }
\vskip 0.6cm
\centerline{\Large\bf for $ CP^{3},CP^{4}$,and $Gr(2,4)$}

\vskip 1.5cm

\centerline{Masao Jinzenji and Yi Sun }
\vskip 0.8cm
\centerline{ \it Department of Phisics, University of Tokyo}
\vskip 0.3cm
\centerline{\it  Bunkyo-ku, Tokyo 113, Japan}
\vskip3cm
\begin{abstract}
 Using the associativity relations of the topological Sigma Models with target
spaces, $CP^3, CP^4$ and $Gr(2,4)$ , we derive recursion relations
of their correlation and evaluate them up to certain order in the expansion
over the instantons. The expansion coeffieients are regarded as the number
 of rational curves in $CP^3, CP^4$ and $Gr(2,4)$ which intersect various
types of submanifolds corresponding to the choice of BRST invariant operators
in the correlation functions.
\end{abstract}

\vfill
\eject

\section{Introduction}
Recently various methods for calculating correlation functions of topological
Sigma Models (A-Model) have been obtained from the analysis of their
structure as the twisted version of $N=2$ super conformal field theory.
One of the most
striking feature is Mirror Symmetry,which emerged from the analysis of
spectrum of 2-types of rings (chiral,chiral) and (chiral,antichiral) of
 Topological Sigma Model on Calabi-Yau mfd and from the speculation that
there are pairs of Calabi-Yau manifold $M$ and $\tilde{M}$ such that
(chiral,chiral) ring of $M$ (resp.$\tilde{M}$) and (chiral,antichiral)
ring of  $\tilde{M}$ (resp.$M$) coincide. Another feature is recently found
by Kontsevich, Manin \cite{km} and Dubrovin \cite{dub} ,which uses the
associativity of
chiral algebra for Topological Sigma Models (coupled with gravity but
on small phase space).
In \cite{km}, the authors defined Gromov-Witten classes as cohomology
classes of moduli space of punctured $CP^{1}$ and showed that associativity
condition
is reduced to the characteristic of boundaries of the moduli space under the
assumption of splitting axiom. Using the method of algebraic geometry numbers
of rational curves in M are counted which pass through  Poincare duals
of cohomology classes corresponding to inserted chiral primary fields. They
showed that for Fano
varieties $M$ whose cohomology are generated by $H^{2}(M)$, the associativity
condition is enough for
the construction of all correlation functions.

With these results available we explicitcy calculate correlation functions of
$CP^{3}$,$CP^{4}$ and $Gr(2,4)$. In the case of $CP^{3}$ and $CP^{4}$ we obtain
a series of overdetermined differential equation for the free energy.
We assume the expansion of the free energy as a generating function of
topological amplitudes and impose ghost number selection rule. Then we pick
a subset of these equations with linear terms and derive recurtion relations
which determine the correlation functions coming from moduli
space of maps of degree $d$ from the ones of lower degree. Then we observe
that all of them are determined and calculate them to several degree d. For
$CP^{3}$,$d=6$,$CP^{4}$,$d=4$,and for $Gr(2,4)$,$d=4$. We checked
compatibility of all the equations in our calculation up to the degree
we calculated.

In section 1, we introduce the  interpretation of correlation functions
as intersection numbers of holomorphic maps from $CP^{1}$ to the target
space $M$ originally derived by Witten \cite{w2} and explain when
 coupled to gravity, they count the number of rational curves passing
through Poincar\'e duals of inserted operators. In section 2, assuming that
these models are twisted version of $N=2$ super conformal field theory
and have stable property of chiral algebra under pertubation of chiral
primaary fields, we set up the calculation of the free energy $F_{M}$ using
associativity
of this algebra. In section 3, we carry out the calculation for each case
$CP^{3}$,$CP^{4}$ and $Gr(2,4)$. Recurtion relations which are
used in determination of amplitudes are written out explicitly. All results
are
collected in tables at the end of this paper.

\section{Meaning of the Correlation Function}
\setcounter{equation}{0}

 In the topological sigma model (A-model) which describes maps from $CP^1$
to the
target space $M$, BRST-closed observables are constructed from elements of
$H^{*}(M)$ {\cite{w1}. We denote the BRST-closed observable
constructed from
$W \in H^{*}(M)$ as ${\cal O}_{W}$. Witten  showed in the pure matter case
\cite{w2}
(without coupling to gravity) topological correlation functions
are given in terms of intersection numbers of holomorphic maps from $CP^{1}$
to $M$ as follows.
\begin{eqnarray}
\langle {\cal O}_{W_{i_1}}(z_{1}){\cal O}_{W_{i_2}}(z_{2}) \cdots
{\cal O}_{W_{i_k}}(z_{k})
\rangle& =& \int_{{\cal M}_{d}(M)} \chi(\nu) \prod_{j=1}^{k} \phi_{j}^{*}
(W_{i_j})\\
 \phi_{j} : {\cal M}_{d}(M)& \mapsto &M\nonumber\\
               f \in {\cal M}_{d}(M) , z_j & \mapsto &f(z_{j}) \in M\nonumber
\;\;\;\;j=1,\cdots,k
\label{a1}
\end{eqnarray}
(${\cal M}_{d}(M)$ is the moduli space of holomorphic maps from $CP^{1}$ to
$M$ of degree d,and $(z_{1},\ldots,z_{k})$ are ``fixed'' distinct points
on $CP^{1}$. Degree $d$ is related to the sum of $dim_{C}(W_{i_j})$ by the
topological selection rule which we will introduce later ).

$\nu$ is the additional degree of freedom which arises when f can be
decomposed as $f = \tilde{f}\circ \alpha $ where $\alpha$ is a map from
$CP^{1}$ to $CP^{1}$ of degree $\frac {d}{j}$ and $\tilde{f}$ a map from
$CP^{1}$ to $M$ of degree $j (d|j)$. But as we will discuss later, we have
to consider $\nu$ only when $M$ is C.Y manifold, i.e. $c_{1}(TM)=0$.

 Since $\phi_{j}^{*}(W_{i_j})$ defines $dim_{C}(W_{i_j})$ form on
${\cal M}_{d}(M)$, in generic case when $\nu$ is trivial, $\langle
{\cal O}_{W_{i_1}} \cdots {\cal O}_{W_{i_k}}\rangle$ doesn't vanish only
when the following conditions are satisfied.
\begin{eqnarray}
 \sum_{j=1}^{k}dim_C(W_{i_j}) & = & dim_{C}({\cal  M}_{d}(M))\nonumber\\
                              & = & dim H^{0}(f^{*}(TM))\nonumber\\
                              & = & dc_1(TM) + dim_{C}(M)
\label{a2}
\end{eqnarray}
In deriving the third line from the second line, we used Rieman-Roch theorem
and assumed $H^{1}(f^{*}(TM))=0$.

If we take $W_{i_j}$ as the form which has a  delta-function support on the
Poincare dual of $W_{i_j}$, $PD(W_{i_j})$, we can interpret
$\phi_{i}^{*}(W_{i_j})$ as the following
constraint on ${\cal M}_{d}(M)$.
\begin{equation}
      f(z_j) \in PD(W_{i_j})
\label{a3}
\end{equation}
We can easily see the above condition imposes $dim_{C}(W_i)$ independent
constraints on ${\cal M}_{d}(M)$ (use count degrees of freedom in the complex
sense). Since we have to use $(dim_{C}(M)-dim_{C}(f(CP^1))-
dim_{C}(PD(W_{i_j}))$ degrees of freedom to let $f(CP^1) \cap PD(W_{i_j})
\ne \emptyset$ and in case $dim_{C}(f(CP^1))=1$, we have to use one further
degree of freedom to let $z_j$ to lie on
$f(CP^{1}) \cap PD(W_{i_j})$.
Condition (\ref{a3}) tells us that by imposing all the constraints
$i=1,\cdots,k$, we have zero degrees of freedom and topological correlation
functions reduce to
\begin{eqnarray}
\lefteqn{\langle {\cal O}_{W_{i_1}}(z_1)\cdots {\cal O}_{W_{i_k}}(z_k)
\rangle_{generic}}\nonumber\\
&& = {}^{\sharp}\{ f: CP^{1}
\stackrel{hol.}{\longrightarrow}  M|f(z_j) \in PD(W_{i_j}),
 j=1,\cdots,k \}
\label{a4}
\end{eqnarray}
 At this point, we consider the case of multiple cover map.
{}From the above argument, multiple cover map $f = \tilde{f}\circ \alpha $
also has to satisfy the condition (\ref{a3}) which restricts the motion
of $f(CP^1)=\tilde{f}(CP^1)$ in the target space M. But since $\tilde{f}$ is
a map of degree j, it has as many as $jc_1(TM)+dim_{C}(M) (<dc_1(TM)+
dim_{C}(M))$ freedom in M and this is imcompatible with (\ref{a3}). Only
when $c_1(TM)=0$ i.e. $M$ is C.Y manifold, compatibility of (\ref{a2}) and
(\ref{a3}) holds in the case of multiple cover map and we have to integrate the
additional $\nu$. Then we conclude that when $c_1(TM)>0$,we can neglect
$\chi(\nu)$ and only consider the generic case.

Next, let us consider what happens if we couple topological gravity to the
above
topological sigma model. Roughly speaking, we have to integrate over moduli
space of $CP^{1}$ with punctures. Since the moduli space of $CP^{1}$ with
$k$-punctures are given by the position of $k$-distinct points on $CP^1$
divided by $SL(2,C)$, which is the internal symmetry group of $CP^{1}$, the
difinition (\ref{a1}) is modified as follows.
\begin{eqnarray}
\langle {\cal O}_{W_{i_1}}\cdots{\cal O}_{W_{i_k}} \rangle& =&
\int _{{\cal M}_{d,0,k}(M)} \prod_{j=1}^{k} \tilde{\phi_{j}}^{*}(W_{i_{j}})\\
 \tilde{\phi_{j}} : {\cal M}_{d,0,k}(M) &\mapsto& M \nonumber\\
  (z_1,z_2,\cdots,z_k,f)/SL(2,C) &\mapsto& (f(z_1),\cdot\cdot\cdot,f(z_k))
\nonumber
\label{a5}
\end{eqnarray}
where the action of $u \in SL(2,C)$ is defined as follows.
\begin{equation}
u\circ(z_1,z_2,\cdots,z_k,f)= (u(z_1),u(z_2),\cdots,u(z_k),(u^{-1})^{*}
f)
\label{a6}
\end{equation}
This action is natural in the sense that the image of $\tilde{\phi_{j}}$
remains invariant under $SL(2,C)$. Main difference between (\ref{a4}) and
(\ref{a6}) is that in the former case, we keep $z_{i}$ ``fixed'' on
$CP^{1}$ but in the latter they move. Then we have
$dim_{C}({\cal M}_{d,0,k}(M))= k-3 + dc_{1}(TM)+dim_{C}(M)$ and modify
(\ref{a2}) as follows.
\begin{eqnarray}
  \sum_{j=1}^{k}dim_{C}(W_{i_j}) = dc_{1}(TM) + dim_{C}(M) +k-3\nonumber\\
 \Longleftrightarrow \sum_{j=1}^{k}dim_{C}(W_{i_j}-1)=
  dc_1(TM) + dim_{C}(M)-3
\label{a7}
\end{eqnarray}
 Integrating over the positions of $k$-punctures the codition (\ref{a3})
changes into
\begin{equation}
   f(CP^{1})\cap PD(W_{i_j})\neq \emptyset .
\label{a8}
\end{equation}
Under the condition (\ref{a7}), $f(CP^{1})\cap PD(W_{i_{j}})$ must be a
finite point set for each $j$ and $z_i$ integration contributes
$(f(CP^{1})\cap PD(W_{i_{j}}))^{\sharp}$ to the correlation function.
Then we have
\begin{eqnarray}
\langle{\cal O}_{W_{i_1}}\cdots {\cal O}_{W_{i_k}}\rangle
= \sum _{f}\prod_{j=1}^{k}(f(CP^{1}) \cap PD(W_{i_{j}}))^{\sharp}\nonumber\\
\{ f: CP^{1} \stackrel{hol.}{\longrightarrow}  M |
 f(CP^{1})\cap PD(W_{i_{j}})\neq \emptyset\quad  j =1,2,\cdots,k \}
\label{a9}
\end{eqnarray}

\section{Set up of the calculation}

\setcounter{equation}{0}

Topological Sigma Model (A-Model) can be constructed as the twisted version
of $N=2$ super conformal field theory \cite{ey}. We can perturb topological
field theory
by adding the terms $\sum_{i} t_{i}{\cal O}_{W_i}$ to the lagrangian and
correlation functions depend on variables $\{t_i\}$ \cite{cv}.
\begin{eqnarray}
\langle {\cal O}_{W_{i_{1}}} {\cal O}_{W_{i_{2}}} \cdots {\cal O}_{W_{i_{k}}}
(t_1,t_2,\ldots,t_{dim(H^{*}(M))})\rangle \nonumber \\
= \int {\cal D}X e ^{-L(X)+\sum_{i} t_{i}{\cal O}_{W_i}} {\cal O}_{W_{i_{1}}}
{\cal O}_{W_{i_{2}}} \cdots {\cal O}_{W_{i_{k}}}
\label{b1}
\end{eqnarray}
where X denotes the field variables of the A-Model. We set $D=dim(H^{*}(M))$.
As the twisted version of N=2 SCFT (coupled with gravity and on small
 phase space ),$\{ {\cal O}_{W_i} \}$ has ring structure which can be
determined from three point correlation functions.
\begin{eqnarray}
    {\cal O}_{W_i}{\cal O}_{W_j}=C^{k}_{ij}(t_1,t_2,\ldots,t_D){\cal O}_{W_k}\\
    \mbox{where}\quad C^{k}_{ij}= C_{ijl}\eta^{lk}\\
     C_{ijl} = \langle {\cal O}_{W_i} {\cal O}_{W_j} {\cal O}_{W_l} (t_1,t_2,
\ldots,t_D) \rangle\\
         \eta^{kl}\eta_{lm} = \delta^{k}_{m}\\
          \eta_{lm} = C_{1lm}(t_1,t_2,\ldots,t_D)
\label{b2}
\end{eqnarray}
In our notation $W_{1}$ corresponds to $1 \in H^{*}(M)$ and we set $W_{2}$ to
the
K\"ahler form of $M$ (in our case where $M$ is $CP^{3},CP^{4}$ or
$Gr(2,4)$, $dim(H^{2}(M))=dim(H^{1,1}(M))=1$,this notation is O.K).
 We assume that $t_{i}$'s are flat coordinates and
$\eta_{lm}$ do not depend on them and determined by classical intersection
number $\int_{M} W_{l} \wedge W_{m}$. Next, we impose associativity
condition on this algebra.(This relation is DWVV eq.)
\begin{eqnarray}
     && ({\cal O}_{W_i}{\cal O}_{W_j}){\cal O}_{W_k}={\cal O}_{W_i}(
{\cal O}_{W_j} {\cal O}_{W_k}) \nonumber\\
&\Longleftrightarrow&  C^{l}_{ij}{\cal O}_{W_l}{\cal O}_{W_k}={\cal O}_{W_i}
C^{m}_{jk}{\cal O}_{W_m}\nonumber\\
&\Longleftrightarrow& C^{l}_{ij}C^{n}_{lk}{\cal O}_{W_n} = C^{m}_{jk}C^{n}_{im}
{\cal O}_{W_n}\nonumber\\
&\Longleftrightarrow& C^{l}_{ij}C^{n}_{lk}=C^{m}_{jk}C^{n}_{im}\nonumber\\
&\Longleftrightarrow& C_{ijm}\eta^{lm}C_{lkn}=C_{jkm}\eta^{lm}C_{imn}
\label{b3}
\end{eqnarray}
And there exists a free energy (prepotential) $F_{M}(t_1,\ldots,t_D)$ which
satisfies following
conditions.
\begin{eqnarray}
   C_{ijk}(t_1,t_2,\ldots,t_D)= \partial_{i}\partial_{j}\partial_{k}F_{M}\\
       (\partial_{i} := \frac{\partial}{\partial t_{i}})
\label{b4}
\end{eqnarray}
 Combining (\ref{b3}) and (\ref{b4}), we obtain a series of partial
differential
equations for $F_{M}$.
\begin{equation}
 \eta^{lm}\partial_{i}\partial_{j}\partial_{m} F_{M}\partial_{l}\partial_{k}
\partial_{n} F_{M} = \eta^{lm}\partial_{j}\partial_{k}\partial_{m} F_{M}
\partial_{i}\partial_{l}\partial_{n} F_{M}
\label{b5}
\end{equation}
 We can also consider prepotential as the generating function of all
topological correlation functions.
\begin{eqnarray}
 F_{M}(t_1,\ldots,t_{D}) =
\sum_{n_1,\cdots,n_{D} \geq 0}\langle {\cal O}^{n_1}_{W_1}
\cdots {\cal O}^{n_{D}}_{W_{D}}\rangle
\frac{t_{1}^{n_{1}}}{n_{1}!}\cdots\frac{t_{D}^{n_{D}}}{n_{D}!}
\label{b6}
\end{eqnarray}
(${\cal O}^{n_{i}}_{W_{i}}$ represents $\overbrace{{\cal O}_{W_i}\cdots
{\cal O}_{W_i}}^{\mbox{$n_{i}$ times}}$ and should not be confused with
${\cal O}_{(W_{i})^{n_i}}$ ).
 At the topological point (i.e., all the $t_{i}$'s are set to zero)
correlation functions become intersection numbers
on moduli spaces of holomorphic maps from $CP^{1}$ (with $k$-- marked points)
to target space $M$.

 Holomorphic maps $f$ are characterized by their degree which equal
the intersection number of $f^{*}(CP^{1})$ with the K\"ahler form
of target space $M$.
 Then $\langle {\cal O}^{n_1}_{W_1}\cdots{\cal O}^{n_D}_{W_{D}}\rangle$
 remains non--zero only when the following topological selection rule
is satisfied.
\begin{eqnarray}
\sum_{i=1}^{D} n_{i}dim_{C}(W_i) = \sum_{i=1}^{D}n_{i} -3 +
dim H^{0}(CP^{1},\phi^{*}(TM)) \nonumber \\
\Longleftrightarrow \sum_{i=1}^{D}n_{i}(dim_{C}(W_{i})-1)= -3 + dc_1(TM)
+ dim_{C}(M)
\label{b7}
\end{eqnarray}
Here d is the degree of holomorphic map and we used Riemann--Roch theorem
in deriving the second line from the first one. When d equals zero, $f$
is  a constant map to the target space and moduli space becomes just the direct
product of target space and moduli space of $CP^{1}$ with
$\displaystyle{\sum_{i=1}^{D} n_i}$ punctures. Then selection rules (\ref{b7})
decomposes to
\begin{equation}
\sum_{i=1}^{D}n_i dim_{C}(W_i) = dim_{C}(M)
\label{b8}
\end{equation}
 and
\begin{equation}
\sum_{i=1}^{D} n_i =3
\label{b9}
\end{equation}

 From (\ref{b9}), we conclude that in $d = 0$ case only $3$-point functions
survive and correlation functions are just classical intersection numbers
$\int _{M}W_{i}\wedge W_{j}\wedge W_{k}$.

{}From the flat metric condition, insertion of $W_{1}$ remains non-zero only
for three point functions from constant maps because one and two point
functions including $W_{1}$ cannot remein nonzero when $c_{1}(M) \geq 1$
and $d \geq 1$ and if we suppose $n (\geq3)$ point functions remain nonzero
in $d \geq 1$ sector, flat metric condition is broken(in three point functions
 in $d \geq 1$ sector, we take into account of the
insertion of operator ${\cal O}_{W_{2}}(dim_{C}(W_2)=1)$ which we will discuss
later).

 With these considerations, expansion of the free energy becomes
\a
F_{M}(t_1,\ldots ,t_D)
=\frac 16 \int_{M}(\sum_{i=1}^{D} t_i W_i )^{3} + \sum_ {d=1}^{\infty}
 \sum _{n_2,\cdots,n_D\geq 0} \frac{{t_{2}}^{n_{2}}}{n_{2}!}\cdots
\frac{{t_{D}}^{n_{D}}}{n_{D}!}\langle {\cal O}^{n_2}_{W_{2}}\cdots
{\cal O}^{n_D}_{W_{D}}\rangle\nonumber\\
 (\sum_{i=2}^{D} n_{i}(dim_{C}(W_i)-1)= -3 + dc_{1}(TM)+dim_C(M))
\label{b10}
\b
 (where the product of the first term of r.h.s. means taking the wedge
product of $H^{*}(M)$).

 Next, we consider the insertion of the operator ${\cal O}_{W_2}$ which
corresponds to the K\"ahler form of the target space . Since
$ codim_{C}(PD(W_{2}))=1$, holomorphic map $f$ of degree $d$
always intersects with it in $d$-points,and the condition
$f(CP^{1})\cap PD(W_{2}) \neq \emptyset$ imposes no constraint. Then from
(\ref{a9}) the
 insertion of ${\cal O}_{W_2}$ results in the multiplication by a factor d,
\begin{equation}
\langle {\cal O}^{n_{2}}_{W_{2}}{\cal O}^{n_{3}}_{W_{3}}
\cdots {\cal O}^{n_{D}}_{W_{D}}\rangle =
d^{n_{2}}\langle {\cal O}^{n_{3}}_{W_{3}}\cdots {\cal O}^{n_{D}}_{W_{D}}
\rangle
\label{b11}
\end{equation}
 Combining (\ref{b10}) and (\ref{b11}) we obtain the following expansion
\a
F_{M}(t_1,\ldots ,t_D)
= \frac 16 \int_{M}(\sum_{i=1}^{D} t_i W_i )^{3} + \sum_{d=1}^{\infty}
 \sum _{n_3,\cdots,n_D\geq 0} \frac{{t_{3}}^{n_{3}}}{n_{3}!}\cdots
\frac{{t_{D}}^{n_{D}}}{n_{D}!}\langle {\cal O}^{n_3}_{W_{3}}\cdots
{\cal O}^{n_D}_{W_{D}}\rangle e^{dt_{2}}\nonumber\\
(\sum_{i=3}^{D}n_{i}(dim_{C}(W_{i})-1)=-3+dc_{1}(TM)+dim_{C}(M))
\label{b12}
\b
 Then by combining (\ref{b5}) and (\ref{b12}), we determine the correlation
functions in the case where target space is $CP^{3},CP^{4}$ and $Gr(2,4)$.

\section{The Calculations}

\setcounter{equation}{0}


The non zero Betti numbers of $CP^3$, $CP^4$ and $Gr(2,4)$ are
\a
         b_{00} = b_{11} = b_{22} = b_{33} = 1
\b
\a
         b_{00} = b_{11} = b_{22} = b_{33} = b_{44}= 1
\b
\a
         b_{00} = b_{11} = b_{33} = b_{44} = 1, \qquad   b_{22}=2
\b
respectively. By means of wedge product we obtain an associative, commutative
ring $H^*(M,Q)$ for each manifold $M$.For $CP^3$, $CP^4$, and $Gr(2,4)$ we have
 the multiplication table as follows.
\clearpage
\begin{table}[h]
\caption{\bf The ring of $CP^3$}
  \begin{center}
    \begin{tabular}{|l|l|l|l|l|}
     \hline
$ $  & $W_1$ & $W_2$ & $W_3$ & $W_4$ \\ \hline
$W_1$ & $W_1$ & $W_2$ & $W_3$ & $W_4$ \\ \hline
$W_2$ & $W_2$ & $W_3$ & $W_4$ & $0$ \\ \hline
$W_3$ & $W_3$ & $W_4$ & $0$ & $0$ \\ \hline
$W_4$ & $W_4$ & $0$ & $0$ & $0$ \\ \hline
\multicolumn{5}{c}{$\scriptstyle{dim_{\cal C}(W_1)=0\; dim_{\cal C}(W_2)=1\;}$}
\\
\multicolumn{5}{c}{$\scriptstyle{dim_{\cal C}(W_3)=2\; dim_{\cal
C}(W_4)=3\;}$}\\
\end{tabular}
      \end{center}
     \end{table}
\begin{table}[h]
\caption{\bf The ring of $CP^4$}
  \begin{center}
    \begin{tabular}{|l|l|l|l|l|l|}
     \hline
$ $  & $W_1$ & $W_2$ & $W_3$ & $W_4$ & $W_5$ \\ \hline
$W_1$ & $W_1$ & $W_2$ & $W_3$ & $W_4$ & $W_5$ \\ \hline
$W_2$ & $W_2$ & $W_3$ & $W_4$ & $W_5$ & $0$ \\ \hline
$W_3$ & $W_3$ & $W_4$ & $W_5$ & $0$ & $0$ \\ \hline
$W_4$ & $W_4$ & $W_5$ & $0$ & $0$ & $0$ \\ \hline
$W_5$ & $W_5$ & $0$ & $0$ & $0$ & $0$\\ \hline
\multicolumn{6}{c}{$\scriptstyle{dim_{\cal C}(W_1)=0\; dim_{\cal
C}(W_2)=1\;}$}\\
\multicolumn{6}{c}{$\scriptstyle{dim_{\cal C}(W_3)=2\; dim_{\cal C}(W_4)=3\;
dim_{\cal C}(W_5)=4\;}$}\\
\end{tabular}
      \end{center}
     \end{table}

\begin{table}[h]
\caption{\bf The ring of $Gr(2,4)$}
  \begin{center}
    \begin{tabular}{|l|l|l|l|l|l|l|}
     \hline
$ $  & $W_1$ & $W_2$ & $W_3$ & $W_4$ & $W_5$ & $W_6$ \\ \hline
$W_1$ & $W_1$ & $W_2$ & $W_3$ & $W_4$ & $W_5$ & $W_6$ \\ \hline
$W_2$ & $W_2$ & $W_3+W_4$ & $W_5$ & $W_5$ & $W_6$ & $0$ \\ \hline
$W_3$ & $W_3$ & $W_5$ & $W_6$ & $0$ & $0$ & $0$ \\ \hline
$W_4$ & $W_4$ & $W_5$ & $0$ & $0$ & $0$ & $0$ \\ \hline
$W_5$ & $W_5$ & $W_6$ & $0$ & $0$ & $0$ & $0$ \\ \hline
$W_6$ & $W_6$ & $0$ & $0$ & $0$ & $0$ & $0$ \\ \hline
\multicolumn{7}{c}{$\scriptstyle{dim_{\cal C}(W_1)=0\; dim_{\cal C}(W_2)=1\;
dim_{\cal C}(W_3)=2\;}$}\\
\multicolumn{7}{c}{$\scriptstyle{dim_{\cal C}(W_4)=2\; dim_{\cal
C}(W_5)=3\;dim_{\cal C}{W_6}=4\;}$}\\
\end{tabular}
      \end{center}
     \end{table}

Dual to each cohomology class is a class of cycles (e.g. for the case of $CP^3$
$W_3$ is dual to a point, $W_2$ is dual to a line, $W_1$ is dual to the
$CP^3$).

As a point intersects the $CP^3$ in a point and a line intersects the plane by
a point. Thus we have for $CP^3$,
\a
<W_1,W_4>=1, <W_2,W_3>=1
\b
For $CP^4$
\a
<W_1,W_5>=1, \qquad <W_2,W_4>=1, \qquad <W_3,W_3>=1
\b
and for Gr(2,4) it becomes
\a
<W_1,W_6>=1, \qquad <W_2,W_5>=1 \\
<W_3,W_3>=1, \qquad <W_4,W_4>=1
\b
All other intersections on generators being zero. The $CP^3$, $CP^4$ ring can
be identified the ring of polynomials in one indeterminate C[x] modulo the
gradient of
\a
W(x)=x^4/4, \qquad W(x)=x^5/5
\b

Ine the case of Grassmanians their cohomology $H^*(Gr,Q)$ can't be generated by
$H^2(Gr,Q)$.
The cohomology ring of Grassmanian $Gr(2,4)$ for instance, can be written as
the singularity ring generated by a single potential\cite{va}
\a
          W(x_i)={\frac 1{5}}x^5_1-x^3_1x_2+x^2_2x_1
\b
Where $x_1$ correspond to $W_2$ and $x_2$ correspond to ${\frac
1{2}}(W_3+W_4)$.



{}From (3.6) one can split $F_M$ into a classical part and instanton correction
part as
\a
      F_M=f_{cl}+f_M
\b
So for $ CP^3$, $CP^4$ and $Gr(2,4)$ we have
\a
        F_{CP^3}= {\frac 1{2}t_1^2t_4} + t_1t_2t_3 + {\frac 1{6}t_2^3} +
f_{CP^3}(t_2,t_3,t_4)
\b\a
        F_{CP^4}= {\frac 1{2}t_1^2t_5} + {\frac 1{2}t_1t_4^2} + {\frac
1{2}t_3^2t_4} + t_1t_2t_5 + f_{CP^4}(t_2,t_3,t_4,t_5)
\b\a
        F_{Gr(2,4)}= {\frac 1{2}t_1^2t_6} + {\frac 1{2}t_1t_3^2} + {\frac
1{2}t_1t_4^2}
+ t_1t_1t_5 + {\frac 1{2}t_2^2t_3} + {\frac 1{2}t_2^2t_4}\0\\
 + f_{Gr(2,4)}(t_2,t_3,t_4,t_5,t_6)
\b
{}From (2.7) the Riemann-Roch theorem tell us
\a
dimH^0(CP^1,f^*(TM))-3=(dimM-3)+dc_1(TM)
\b
Once we specify the target space , we know its frist Chern class, then the
above formula give the dimension of its moduli space.
In case of $CP^3$
$c_1(TCP^3)=4$
so,
\a
dimH^0(CP^1,f^*(TCP^3))-3=4d
\b
For $CP^4$ and $Gr(2,4)$, the frist Chern class are
\a
c_1(CP^4)=5, \qquad  c_1(TGr(2,4))=4
\b
Thus
\a
dimH^0(CP^1,f^*(TCP^4))-3=5d+1\\  \noal
dimH^0(CP^1,f^*(TGr(2,4)))-3=4d+1
\b
{}From (\ref{b12}) we can expond $f_M$ further as follows
\a
f_{CP^3}=\sum_{d=1}^{\infty}\sum_{n_3+2n_4=4d}{\frac{<{\cal O}_{W_3}^{n_3}
 {\cal O}_{W_4}^{n_4}>}{n_{3}! n_{4}!}}t_3^{n_3}t_4^{n_4}e^{dt_2}\0\\\noal
=\sum_{d=1}^{\infty}\sum_{n_4}{\frac{<{\cal O}_{W_3}^{4d-2n_4}
 {\cal O}_{W_4}^{n_4}>}{(4d-2n_{4})! n_{4}!}}t_3^{4d-2n_4}t_4^{n_4}e^{dt_2}
\label{d}
\b

\a
f_{CP^4}=\sum_{d=1}^{\infty}\sum_{n_3+2n_4+3n_5=5d+1}{\frac{<{\cal
O}_{W_3}^{n_3}
 {\cal O}_{W_4}^{n_4}{\cal O}_{W_5}^{n_5}>}{n_{3}!
n_{4}!n_{5}!}}t_3^{n_3}t_4^{n_4}t_5^{n_5}e^{dt_2}\0\\\noal
=\sum_{d=1}^{\infty}\sum_{n_4,n_5}{\frac{<{\cal O}_{W_3}^{5d-2n_4-3n_5+1}
 {\cal O}_{W_4}^{n_4}{\cal O}_{W_5}^{n_5}>}{(5d-2n_{4}-3n_{5}+1)!
n_{4}!n_{5}!}}t_3^{5d-2n_4-3n_{5}+1}t_4^{n_4}t_5^{n_5}e^{dt_2}\label{e}
\b

\a
f_{Gr(2,4)}=\sum_{d=1}^{\infty}\sum_{n_3+n_4+2n_5+3n_6=4d+1}{\frac{<{\cal
O}_{W_3}^{n_3}
 {\cal O}_{W_4}^{n_4}{\cal O}_{W_5}^{n_5}{\cal
O}_{W_6}^{n_6}>}{n_{3}!n_{4}!n_{5}!n_6!}}t_3^{n_3}t_4^{n_4}t_5^{n_5}t_6^{n_6}e^{dt_2}\0\\\noal
=\sum_{d=1}^{\infty}\sum_{n_4,n_5,n_6}{\frac{<{\cal O}_{W_3}^{4d-n_4-2n_5-3n_6+
1}
 {\cal O}_{W_4}^{n_4}{\cal O}_{W_5}^{n_5}{\cal
O}_{W_6}^{n_6}>}{(4d-n_{4}-2n_{5}-3n_6+1)! n_{4}!n_{5}!}}
t_3^{4d-n_4-2n_{5}-3n_6+1}t_4^{n_4}t_5^{n_5}t_6^{n_6}e^{dt_2}\label{f}
\b

Then, we abbreviate the notion in the following calculation as
\a
{<{\cal O}_{W_3}^{4d-2n_4}{\cal O}_{W_4}^{n_4}>}_{CP^3}=N_{n_4}^d\\\noal
{<{\cal O}_{W_3}^{5d-2n_4-3n_5+1}{\cal O}_{W_4}^{n_4}{\cal
O}_{W_5}^{n_5}>}_{CP^4}=N_{n_4,n_5}^d\\\noal
{<{\cal O}_{W_3}^{4d-n_4-2n_5-3n_6+1}{\cal O}_{W_4}^{n_4}{\cal
O}_{W_5}^{n_5}{\cal O}_{W_6}^{n_6}>}_{Gr(2,4)}
=N_{n_4,n_5,n_6}^d
\b
We let $t_2=x,t_3=y,t_4=z$ for $CP^3$,$t_2=w,t_3=x,t_4=y,t_5=z$ for $CP^4$ and
$t_2=v,t_3=w,t_4=x,t_5=y,
t_6=z$ for $Gr(2,4)$.
A deformation of the multiplication table (table 1, table 2, table 3) become
the fusion rules
for the quantum cohomology ring with ${\cal O}_{W_i}$'s substituted for
$t$'s as
\a
{\cal O}_{W_i} \circ {\cal O}_{W_j} =
{\partial}_i{\partial}_j{\partial}_lF_M\eta^{lk}{\cal O}_{W_k}
\b
The structure constants of the quantum cohomology obey the so called WDVV
equation which satisfying the requirements[3]

(i)commutativity

(ii)associativity

(iii)existence of a unit ${\cal O}_{W_1}$

Commutativity follows from the definition, while condition(3.6)
(equivalently(3.17)
expresses that ${\cal O}_{W_1}$
plays the role of unit. The crucial assumption is the associativity which
imposes strong conditions on $f_M$. Now let us introduce
some more notations, by $f_{M,xyz}$ we mean
${\partial}_x{\partial}_y{\partial}_zf_M $. In the following we will simply
 omit the index ``M'', and just denote it
as $f_{xyz}$.


The quantum ring of $CP^3$ is
\ai
{\cal O}_{W_2}{\cal O}_{W_2}&=&f_{xxz}{\cal O}_{W_1}+f_{xxy}{\cal
O}_{W_2}+(f_{xxx}+1){\cal O}_{W_3},\\\noal
{\cal O}_{W_2}{\cal O}_{W_3}&=&f_{xyz}{\cal O}_{W_1}+f_{xyy}{\cal
O}_{W_2}+(f_{xxy}){\cal O}_{W_3}+{\cal O}_{W_4},\\\noal
{\cal O}_{W_2}{\cal O}_{W_4}&=&f_{xzz}{\cal O}_{W_1}+f_{xyz}{\cal
O}_{W_2}+(f_{xxz}){\cal O}_{W_3},\\\noal
{\cal O}_{W_3}{\cal O}_{W_3}&=&f_{yyz}{\cal O}_{W_1}+f_{yyy}{\cal
O}_{W_2}+(f_{xyy}){\cal O}_{W_3},\\\noal
{\cal O}_{W_3}{\cal O}_{W_4}&=&f_{yzz}{\cal O}_{W_1}+f_{yyz}{\cal
O}_{W_2}+(f_{xyz}){\cal O}_{W_3},\\\noal
{\cal O}_{W_4}{\cal O}_{W_4}&=&f_{zzz}{\cal O}_{W_1}+f_{yzz}{\cal
O}_{W_2}+(f_{xzz}){\cal O}_{W_3}.
\bj
The quantum ring of $CP^4$ is
\ai
{\cal O}_{W_2}{\cal O}_{W_2}&=&f_{wwz}{\cal O}_{W_1}+f_{wwy}{\cal
O}_{W_2}+f_{wwx}{\cal O}_{W_3}+(f_{www}+1){\cal O}_{W_4},\\\noal
{\cal O}_{W_2}{\cal O}_{W_3}&=&f_{wxz}{\cal O}_{W_1}+f_{wxy}{\cal
O}_{W_2}+f_{wxx}{\cal O}_{W_3}+f_{wwx}{\cal O}_{W_4},\\\noal
{\cal O}_{W_2}{\cal O}_{W_4}&=&f_{wyz}{\cal O}_{W_1}+f_{wyy}{\cal
O}_{W_2}+f_{wxy}{\cal O}_{W_3}+f_{wwy}{\cal O}_{W_4}+{\cal O}_{W_5},\\\noal
{\cal O}_{W_2}{\cal O}_{W_5}&=&f_{wzz}{\cal O}_{W_1}+f_{wyz}{\cal
O}_{W_2}+f_{wxz}{\cal O}_{W_3}+f_{wwz}{\cal O}_{W_4},\\\noal
{\cal O}_{W_3}{\cal O}_{W_3}&=&f_{xxz}{\cal O}_{W_1}+f_{xxy}{\cal
O}_{W_2}+f_{xxx}{\cal O}_{W_3}+f_{wxx}{\cal O}_{W_4}+{\cal O}_{W_5},\\\noal
{\cal O}_{W_3}{\cal O}_{W_4}&=&f_{xyz}{\cal O}_{W_1}+f_{xyy}{\cal
O}_{W_2}+f_{xxy}{\cal O}_{W_3}+f_{wxy}{\cal O}_{W_4},\\\noal
{\cal O}_{W_3}{\cal O}_{W_5}&=&f_{xzz}{\cal O}_{W_1}+f_{xyz}{\cal
O}_{W_2}+f_{xxz}{\cal O}_{W_3}+f_{wxz}{\cal O}_{W_4},\\\noal
{\cal O}_{W_4}{\cal O}_{W_4}&=&f_{yyz}{\cal O}_{W_1}+f_{yyy}{\cal
O}_{W_2}+f_{xyy}{\cal O}_{W_3}+f_{wyy}{\cal O}_{W_4},\\\noal
{\cal O}_{W_4}{\cal O}_{W_5}&=&f_{yzz}{\cal O}_{W_1}+f_{yyz}{\cal
O}_{W_2}+f_{xyz}{\cal O}_{W_3}+f_{wyz}{\cal O}_{W_4},\\\noal
{\cal O}_{W_5}{\cal O}_{W_5}&=&f_{zzz}{\cal O}_{W_1}+f_{yzz}{\cal
O}_{W_2}+f_{xzz}{\cal O}_{W_3}+f_{wzz}{\cal O}_{W_4}.
\bj
The quantum ring of $Gr(2,4)$ is
\ai
{\cal O}_{W_2}{\cal O}_{W_2}&=&f_{vvz}{\cal O}_{W_1}+f_{vvy}{\cal
O}_{W_2}+(f_{vvw}+1){\cal O}_{W_3}\0\\\noal
&+&(f_{vvx}+1){\cal O}_{W_4}+f_{vvv}{\cal O}_{W_5},\\\noal
{\cal O}_{W_2}{\cal O}_{W_3}&=&f_{vwz}{\cal O}_{W_1}+f_{vwy}{\cal
O}_{W_2}+f_{vww}{\cal O}_{W_3}\0\\\noal
&+&f_{vwx}{\cal O}_{W_4}+(f_{vvw}+1){\cal O}_{W_5},\\\noal
{\cal O}_{W_2}{\cal O}_{W_4}&=&f_{vxz}{\cal O}_{W_1}+f_{vxy}{\cal
O}_{W_2}+f_{vwx}{\cal O}_{W_3}+f_{vxx}{\cal O}_{W_4}+(f_{vvx}+1){\cal
O}_{W_5},\\\noal
{\cal O}_{W_2}{\cal O}_{W_5}&=&f_{vyz}{\cal O}_{W_1}+f_{vyy}{\cal
O}_{W_2}+f_{vwy}{\cal O}_{W_3}+f_{vxy}{\cal O}_{W_4}+f_{vvy}{\cal
O}_{W_5},\\\noal
{\cal O}_{W_2}{\cal O}_{W_6}&=&f_{vzz}{\cal O}_{W_1}+f_{vyz}{\cal
O}_{W_2}+f_{vwz}{\cal O}_{W_3}+f_{vxz}{\cal O}_{W_4}+f_{vvz}{\cal
O}_{W_5},\\\noal
{\cal O}_{W_3}{\cal O}_{W_3}&=&f_{wwz}{\cal O}_{W_1}+f_{wwy}{\cal
O}_{W_2}+f_{www}{\cal O}_{W_3}\0\\\noal
&+&f_{wwx}{\cal O}_{W_4}+f_{vww}{\cal O}_{W_5} +{\cal O}_{W_6},\\\noal
{\cal O}_{W_3}{\cal O}_{W_4}&=&f_{wxz}{\cal O}_{W_1}+f_{wxy}{\cal
O}_{W_2}+f_{wwx}{\cal O}_{W_3}+f_{wxx}{\cal O}_{W_4}+f_{vwx}{\cal
O}_{W_5},\\\noal
{\cal O}_{W_3}{\cal O}_{W_5}&=&f_{wyz}{\cal O}_{W_1}+f_{wyy}{\cal
O}_{W_2}+f_{wwy}{\cal O}_{W_3}+f_{wxy}{\cal O}_{W_4}+f_{vwy}{\cal
O}_{W_5},\\\noal
{\cal O}_{W_3}{\cal O}_{W_6}&=&f_{wzz}{\cal O}_{W_1}+f_{wyz}{\cal
O}_{W_2}+f_{wwz}{\cal O}_{W_3}+f_{wxz}{\cal O}_{W_4}+f_{vwz}{\cal
O}_{W_5},\\\noal
{\cal O}_{W_4}{\cal O}_{W_4}&=&f_{xxz}{\cal O}_{W_1}+f_{xxy}{\cal
O}_{W_2}+f_{wxx}{\cal O}_{W_3}\0\\\noal
&+&f_{xxx}{\cal O}_{W_4}+f_{vxx}{\cal O}_{W_5} +{\cal O}_{W_6},\\\noal
{\cal O}_{W_4}{\cal O}_{W_5}&=&f_{xyz}{\cal O}_{W_1}+f_{xyy}{\cal
O}_{W_2}+f_{wxy}{\cal O}_{W_3}+f_{xxy}{\cal O}_{W_4}+f_{vxy}{\cal
O}_{W_5},\\\noal
{\cal O}_{W_4}{\cal O}_{W_6}&=&f_{xzz}{\cal O}_{W_1}+f_{xyz}{\cal
O}_{W_2}+f_{wxz}{\cal O}_{W_3}+f_{xxz}{\cal O}_{W_4}+f_{vxz}{\cal
O}_{W_5},\\\noal
{\cal O}_{W_5}{\cal O}_{W_5}&=&f_{yyz}{\cal O}_{W_1}+f_{yyy}{\cal
O}_{W_2}+f_{wyy}{\cal O}_{W_3}+f_{xyy}{\cal O}_{W_4}+f_{vyy}{\cal
O}_{W_5},\\\noal
{\cal O}_{W_5}{\cal O}_{W_6}&=&f_{yzz}{\cal O}_{W_1}+f_{yyz}{\cal
O}_{W_2}+f_{wyz}{\cal O}_{W_3}+f_{xyz}{\cal O}_{W_4}+f_{}vyz{\cal
O}_{W_5},\\\noal
{\cal O}_{W_6}{\cal O}_{W_6}&=&f_{zzz}{\cal O}_{W_1}+f_{yzz}{\cal
O}_{W_2}+f_{wzz}{\cal O}_{W_3}+f_{xzz}{\cal O}_{W_4}+f_{vzz}{\cal O}_{W_5}.
\bj
Associativity condition (3.7) implies  the free energy of $CP^3$ must satifying
the following constraint equation
\ai
&&-2f_{xyz}-f_{xyy}f_{xxy}+f_{yyy}f_{xxx}=0,\\ \noal
&&-f_{xzz}-f_{xyy}f_{xxz}+f_{yyz}f_{xxx}=0,\\\noal
&&f_{yzz}-f_{xxz}f_{yyy}+f_{xxy}f_{yyz}=0,\\\noal
&&-2f_{xyz}f_{xxz}+f_{xzz}f_{xxy}+f_{yzz}f_{xxx}=0,\\\noal
&&f_{zzz}-f_{xyz}^2+f_{xzz}f_{xyy}-f_{yyz}f_{xxz}+f_{yzz}f_{xxy}=0,\\\noal
&&f_{yyy}f_{xzz}-2f_{yyz}f_{xyz}+f_{yzz}f_{xyy}=0.
\bj

For $CP^4$ there are 17 independent constraint equations. We just write down
five of them which are enough to determine the correlation functions of $CP^4$
\ai
&&-f_{wwz}-f_{wwy}f_{wxx}+f_{wxx}^2+2f_{wwx}f_{xxx}-f_{www}f_{xxy}=0, \\\noal
&&f_{wxy}^2 + f_{wwy}f_{wyy}+2f_{wyz}-f_{wwx}f_{xyy}-f_{www}f_{yyy}=0, \\\noal
&&f_{wxy}f_{wxz}+f_{wwz}f_{wyy}+f_{wzz}-f_{wwx}f_{xyz}-f_{www}f_{yyz}=0,\\\noal
&&-f_{wxz}f_{xyy}+f_{wxy}f_{xyz}-f_{wwz}f_{yyy}+f_{wwy}f_{yyz}+f_{yzz}=0,\\\noal
&&f_{wxy}f_{xxy}+f_{wwy}f_{xyy}-f_{wxx}f_{xyy}+f_{xyz}-f_{wwx}f_{yyy}=0.
\bj

For the case of $Gr(2,4)$ there are fifty-six independent equations. We also
write down nine of them that determine the corelation functions of $Gr(2,4)$
\ai
&& -f_{vvz}-f_{vvy}f_{vww}+f_{vww}^2+f_{vwx}^2+2f_{vvw}f_{vwy}\0\\
&& -f_{vvw}f_{www}-f_{vvx}f_{wwx}-f_{vvv}f_{wwy}=0,\\\noal
&& -f_{vvz}-f_{vvy}f_{vxx}+f_{vwx}^2+f_{vxx}^2+2f_{vvx}f_{vxy} \0\\
&& -f_{vvw}f_{wxx}-f_{vvx}f_{xxx}-f_{vvv}f_{xxy}=0,\\\noal
&& -f_{vxz}-f_{vww}f_{vxy}+f_{vwx}f_{vwy}-f_{vwx}f_{www}+f_{vww}f_{wwx} \0\\
&& -f_{vxx}f_{wwx}-f_{vvx}f_{wwy}+f_{vwx}f_{wxx}+f_{vvw}f_{wxy}=0,\\\noal
&& -f_{wwz}-f_{xxz}-f_{vxx}f_{wwy}+f_{wwx}^2+f_{wxx}^2+2f_{vwx}f_{wxy} \0\\
&& -f_{www}f_{wxx}-f_{wwx}f_{xxx}-f_{vww}f_{xxy}=0,\\\noal
&&-f_{xyz}-f_{vxy}f_{wwy}+f_{wwx}f_{wwy}+f_{vwy}f_{wxy}+f_{wxx}f_{wxy} \0\\
&&-f_{www}f_{wxy}+f_{vwx}f_{wyy}-f_{wwx}f_{xxy}-f_{vww}f_{xyy}=0,\\\noal
&&-f_{xzz}-f_{vxz}f_{wwy}+f_{wwx}f_{wwz}+f_{vwz}f_{wxy}+f_{wxx}f_{wxz} \0\\
&& -f_{www}f_{wxz}+f_{vwx}f_{wyz}-f_{wwx}f_{xxz}-f_{vww}f_{xyz}=0,\\\noal
&& f_{wxy}+f_{vwy}f_{wwy}+f_{vxy}f_{wxy}+f_{vvy}f_{wyy} \0\\
&& -f_{vww}f_{wyy}-f_{vwx}f_{xyy}-f_{vvw}f_{yyy}=0,\\\noal
&&f_{wxz}-f_{vwy}f_{vxy}+f_{vwx}f_{vyy}+f_{vwy}f_{wwx}+f_{vxy}f_{wxx} \0\\
&& -f_{vwx}f_{wwy}+f_{vvy}f_{wxy}-f_{vxx}f_{wxy}-f_{vvx}f_{wyy}=0,\\\noal
&& -f_{yyz}+f_{vwz}f_{wyy}-f_{vwy}f_{wyz}+f_{vxz}f_{xyy} \0\\
&& +f_{vvz}f_{yyy}-f_{vxy}f_{xyz}-f_{vvy}f_{yyz}=0.
\bj

Substituting the free energy (\ref{d}--\ref{f}) into (4.29),(4.30)
and (4.31) one obtains the recursion relations of correlation functions. For
$CP^3$ one has

\ai
&&2dN^d_{m+1}-N^d_{m}=\sum_{\stackrel{f+g=d} {n+n'=m}}\left(\begin{array}{c}
          m \\ n\end{array}\right)\0\\\noal
          &&\bigg[ -\left(\begin{array}{c}
          4d-2m-3 \\ 4f-2n-2 \end{array}\right)
fN^f_{n} N^g_{n'}
    +\left(\begin{array}{c}
          4d-2m-3 \\ 4f-2n-3 \end{array}\right)
g^3N^f_{n} N^g_{n'}     \bigg],\\\noal\noal
&&dN^d_{m+2}-N^d_{m+1}=\sum_{\stackrel{f+g=d} {n+n'=m}}\left(\begin{array}{c}
          m \\ n\end{array}\right)\0\\\noal
          &&\bigg[ \left(\begin{array}{c}
          4d-2m-4 \\ 4f-2n-4 \end{array}\right)
g^3N^f_{n+1} N^g_{n'}
    -\left(\begin{array}{c}
          4d-2m-4 \\ 4f-2n-2 \end{array}\right)
fg^2N^f_{n} N^g_{n'+1}     \bigg],\\\noal\noal
&&N^d_{m+2}=\sum_{\stackrel{f+g=d} {n+n'=m}}\left(\begin{array}{c}
          m \\ n\end{array}\right)\0\\
          &&\bigg[ \left(\begin{array}{c}
          4d-2m-5 \\ 4f-2n-2 \end{array}\right)
f^2N^f_{n+1} N^g_{n'}
    -\left(\begin{array}{c}
          4d-2m-5 \\ 4f-2n-1 \end{array}\right)
f^2N^f_{n} N^g_{n'+1}     \bigg].
\bj
For the case of $CP^4$ the recursion relations read as follows
\ai
&&d^2 N^d_{m_1,m_2+1}-2dN^d_{m_1,m_2}+N^d_{m_1,m_2}\0\\  \noal
&&=\sum_{\stackrel{f+g=d} {n_1+n_1'=m_1, n_2+n_2'=m_2}}\left(\begin{array}{c}
          m_1 \\ n_1 \end{array}\right)
\left(\begin{array}{c}
          m_2 \\ n_2 \end{array}\right)\0\\
          &&\bigg[ -\left(\begin{array}{c}
          5d-2m_1-3m_2-2 \\ 5f-2n_1-3n_2-1 \end{array}\right)
f^2gN^f_{n_1+1,n_2} N^g_{n_1',n_2'}     \0\\
&&  +\left(\begin{array}{c}
          5d-2m_2-3m_3-2 \\ 5f-2n_1-3n_2-1 \end{array}\right)
fgN^f_{n_1,n_2} N^g_{n_1',n_2'}     \0\\
&& +2\left(\begin{array}{c}
          5d-2m_1-3m_2-2 \\ 5f-2n_1-3n_2 \end{array}\right)
f^2gN^f_{n_1,n_2} N^g_{n_1'+1,n_2'}     \0\\
&& -\left(\begin{array}{c}
          5d-2m_1-3m_2-2 \\ 5f-2n_1-3n_2 \end{array}\right)
f^2N^f_{n_1,n_2} N^g_{n_1',n_2'}     \0\\
&& -\left(\begin{array}{c}
          5d-2m_1-3m_2-2 \\ 5f-2n_1-3n_2+1 \end{array}\right)
f^3N^f_{n_1,n_2} N^g_{n_1',n_2'+1}     \bigg],\\\noal\noal
&&N^d_{m_1+1,m_2+1}-dN^d_{m_1,m_2+2}\0\\  \noal
&&=\sum_{\stackrel{f+g=d} {n_1+n_1'=m_1, n_2+n_2'=m_2}}\left(\begin{array}{c}
          m_1 \\ n_1 \end{array}\right)
\left(\begin{array}{c}
          m_2 \\ n_2 \end{array}\right)\0\\
          &&\bigg[ \left(\begin{array}{c}
          5d-2m_1-3m_2-5 \\ 5f-2n_1-3n_2-2 \end{array}\right)
fgN^f_{n_1+1,n_2} N^g_{n_1',n_2'+1}     \0\\
&&  +\left(\begin{array}{c}
          5d-2m_2-3m_3-5 \\ 5f-2n_1-3n_2-2 \end{array}\right)
f^2gN^f_{n_1,n_2+1} N^g_{n_1'+2,n_2'}     \0\\
&& -\left(\begin{array}{c}
          5d-2m_1-3m_2-5 \\ 5f-2n_1-3n_2 \end{array}\right)
f^2N^f_{n_1,n_2} N^g_{n_1'+1,n_2'+1}     \0\\
&& -\left(\begin{array}{c}
          5d-2m_1-3m_2-5 \\ 5f-2n_1-3n_2+1 \end{array}\right)
f^3N^f_{n_1,n_2} N^g_{n_1'+2,n_2'+1}        \bigg],\\\noal\noal
&&N^d_{m_1+2,m_2}-2dN^d_{m_1+1,m_2+1}\0\\  \noal
&&=\sum_{\stackrel{f+g=d} {n_1+n_1'=m_1, n_2+n_2'=m_2}}\left(\begin{array}{c}
          m_1 \\ n_1 \end{array}\right)
\left(\begin{array}{c}
          m_2 \\ n_2 \end{array}\right)\0\\\noal
          &&\bigg[ \left(\begin{array}{c}
          5d-2m_1-3m_2-4 \\ 5f-2n_1-3n_2-2 \end{array}\right)
fgN^f_{n_1+1,n_2} N^g_{n_1'+1,n_2'}     \0\\\noal
&&  +\left(\begin{array}{c}
          5d-2m_2-3m_3-4 \\ 5f-2n_1-3n_2-1 \end{array}\right)
f^2gN^f_{n_1+1,n_2} N^g_{n_1'+2,n_2'}     \0\\\noal
&& -\left(\begin{array}{c}
          5d-2m_1-3m_2-4 \\ 5f-2n_1-3n_2 \end{array}\right)
f^2N^f_{n_1,n_2} N^g_{n_1'+2,n_2'}     \0\\\noal
&& -\left(\begin{array}{c}
          5d-2m_1-3m_2-4 \\ 5f-2n_1-3n_2+1 \end{array}\right)
f^3N^f_{n_1,n_2} N^g_{n_1'+3,n_2'}        \bigg],\\\noal\noal
&&N^d_{m_1+1,m_2+2}\0\\  \noal
&&=\sum_{\stackrel{f+g=d} {n_1+n_1'=m_1, n_2+n_2'=m_2}}\left(\begin{array}{c}
          m_1 \\ n_1 \end{array}\right)
\left(\begin{array}{c}
          m_2 \\ n_2 \end{array}\right)\0\\\noal
          &&\bigg[ \left(\begin{array}{c}
          5d-2m_1-3m_2-7 \\ 5f-2n_1-3n_2-3 \end{array}\right)
fN^f_{n_1,n_2+1} N^g_{n_1'+2,n_2'}     \0\\\noal
&&  -\left(\begin{array}{c}
          5d-2m_2-3m_3-7 \\ 5f-2n_1-3n_2-2 \end{array}\right)
fN^f_{n_1+1,n_2} N^g_{n_1'+1,n_2'+1}     \0\\\noal
&& +\left(\begin{array}{c}
          5d-2m_1-3m_2-7 \\ 5f-2n_1-3n_2 \end{array}\right)
f^2N^f_{n_1,n_2+1} N^g_{n_1'+3,n_2'}     \0\\\noal
&& -\left(\begin{array}{c}
          5d-2m_1-3m_2-7 \\ 5f-2n_1-3n_2-1 \end{array}\right)
f^2N^f_{n_1+1,n_2} N^g_{n_1'+2,n_2'+1}        \bigg].
\bj
For the case of $Gr(2,4)$ the recursion relation of becomes
\ai
&&N^d_{m_1+3,m_2}-N^d_{m_1+1,m_2+1}\0\\  \noal
&&=\sum_{\stackrel{f+g=d} {n_1+n_1'=m_1, n_2+n_2'=m_2}}\left(\begin{array}{c}
          m_1 \\ n_1 \end{array}\right)
\left(\begin{array}{c}
          m_2 \\ n_2 \end{array}\right)\0\\\noal
          &&\bigg[ \left(\begin{array}{c}
          5d-2m_1-3m_2-5 \\ 5f-2n_1-3n_2-2 \end{array}\right)
fN^f_{n_1+1,n_2} N^g_{n_1'+1,n_2'}     \0\\\noal
&&  +\left(\begin{array}{c}
          5d-2m_2-3m_3-5 \\ 5f-2n_1-3n_2-1 \end{array}\right)
f^2N^f_{n_1+1,n_2} N^g_{n_1'+2,n_2'}     \0\\\noal
&& -\left(\begin{array}{c}
          5d-2m_1-3m_2-5 \\ 5f-2n_1-3n_2-1 \end{array}\right)
fN^f_{n_1,n_2} N^g_{n_1'+2,n_2'}     \0\\\noal
&& -\left(\begin{array}{c}
          5d-2m_1-3m_2-5 \\ 5f-2n_1-3n_2 \end{array}\right)
f^2N^f_{n_1,n_2} N^g_{n_1'+3,n_2'}        \bigg],\\\noal\noal\noal
&& d^2 N^d_{m_1,m_2,m_3+1}-2d
N^d_{m_1,m_2+1,m_3}+N^d_{m_1,m_2,m_3}+N^d_{m_1+1,m_2,m_3}  \0\\
&&=\sum_{\stackrel{f+g=d, n_1+n_1'=m_1} {n_2+n_2'=m_2,
n_3+n_3'=m_3}}\left(\begin{array}{c}
          m_1 \\ n_1 \end{array}\right)
\left(\begin{array}{c}
          m_2 \\ n_2 \end{array}\right)
\left(\begin{array}{c}
          m_3 \\ n_3 \end{array}\right)
\0\\
&&\bigg[ - \left(\begin{array}{c}
          4d-m_1-2m_2-3m_3-2 \\ 4f-n_1-2n_2-3n_3-1 \end{array}\right)
f^2g N^f_{n_1,n_2+1,n_3} N^g_{n_1',n_2',n_3'}     \0\\\noal
&&  + \left(\begin{array}{c}
          4d-m_1-2m_2-3m_3-2 \\ 4f-n_1-2n_2-3n_3-1 \end{array}\right)
fg N^f_{n_1,n_2,n_3} N^g_{n_1',n_2',n_3'}     \0\\\noal
&&  + \left(\begin{array}{c}
          4d-m_1-2m_2-3m_3-2 \\ 4f-n_1-2n_2-3n_3-1 \end{array}\right)
fg N^f_{n_1+1,n_2,n_3} N^g_{n_1'+1,n_2',n_3'}     \0\\\noal
&&  + 2 \left(\begin{array}{c}
          4d-m_1-2m_2-3m_3-2 \\ 4f-n_1-2n_2-3n_3 \end{array}\right)
f^2g N^f_{n_1,n_2,n_3} N^g_{n_1',n_2'+1,n_3'}     \0\\\noal
&&   - \left(\begin{array}{c}
          4d-m_1-2m_2-3m_3-2 \\ 4f-n_1-2n_2-3n_3 \end{array}\right)
f^2 N^f_{n_1,n_2,n_3} N^g_{n_1',n_2',n_3'}     \0\\\noal
&&\  - \left(\begin{array}{c}
          4d-m_1-2m_2-3m_3-2 \\ 4f-n_1-2n_2-3n_3\end{array}\right)
f^2 N^f_{n_1+1,n_2,n_3} N^g_{n_1'+1,n_2',n_3'}     \0\\\noal
&&  - \left(\begin{array}{c}
          4d-m_1-2m_2-3m_3-2 \\ 4f-n_1-2n_2-3n_3+1 \end{array}\right)
f^3 N^f_{n_1,n_2,n_3} N^g_{n_1',n_2'+1,n_3'}  \bigg],\\\noal\noal\noal
&& d^2 N^d_{m_1,m_2,m_3+1}-2d
N^d_{m_1+1,m_2+1,m_3}+N^d_{m_1+2,m_2,m_3}+N^d_{m_1+3,m_2,m_3}\0\\  \noal
&&=\sum_{\stackrel{f+g=d, n_1+n_1'=m_1} {n_2+n_2'=m_2,
n_3+n_3'=m_3}}\left(\begin{array}{c}
          m_1 \\ n_1 \end{array}\right)
\left(\begin{array}{c}
          m_2 \\ n_2 \end{array}\right)
\left(\begin{array}{c}
          m_3 \\ n_3 \end{array}\right)\0\\\noal
&&\bigg[ \left(\begin{array}{c}
          4d-m_1-2m_2-3m_3-2 \\ 4f-n_1-2n_2-3n_3-1 \end{array}\right)
fg N^f_{n_1+1,n_2,n_3} N^g_{n_1'+1,n_2',n_3'}     \0\\\noal
&&  - \left(\begin{array}{c}
          4d-m_1-2m_2-3m_3-2 \\ 4f-n_1-2n_2-3n_3-1 \end{array}\right)
f^2g N^f_{n_1,n_2+1,n_3} N^g_{n_1'+2,n_2',n_3'}     \0\\\noal
&&  + \left(\begin{array}{c}
          4d-m_1-2m_2-3m_3-2 \\ 4f-n_1-2n_2-3n_3-1 \end{array}\right)
fg N^f_{n_1+2,n_2,n_3} N^g_{n_1'+2,n_2',n_3'}     \0\\\noal
&&  + 2 \left(\begin{array}{c}
          4d-m_1-2m_2-3m_3-2 \\ 4f-n_1-2n_2-3n_3 \end{array}\right)
f^2g N^f_{n_1+1,n_2,n_3} N^g_{n_1'+1,n_2'+1,n_3'}     \0\\\noal
&&   - \left(\begin{array}{c}
          4d-m_1-2m_2-3m_3-2 \\ 4f-n_1-2n_2-3n_3 \end{array}\right)
f^2 N^f_{n_1,n_2,n_3} N^g_{n_1'+2,n_2',n_3'}     \0\\\noal
&&\  - \left(\begin{array}{c}
          4d-m_1-2m_2-3m_3-2 \\ 4f-n_1-2n_2-3n_3\end{array}\right)
f^2 N^f_{n_1+1,n_2,n_3} N^g_{n_1'+3,n_2',n_3'}     \0\\\noal
&&  - \left(\begin{array}{c}
          4d-m_1-2m_2-3m_3-2 \\ 4f-n_1-2n_2-3n_3+1 \end{array}\right)
f^3 N^f_{n_1,n_2,n_3} N^g_{n_1'+2,n_2'+1,n_3'}    \bigg],\\\noal\noal\noal
&& dN^d_{m_1+1,m_2,m_3+1}+N^d_{m_1,m_2+1,m_3}-N^d_{m_1+1,m_2+1,m_3}\0\\  \noal
&&=\sum_{\stackrel{f+g=d, n_1+n_1'=m_1} {n_2+n_2'=m_2,
n_3+n_3'=m_3}}\left(\begin{array}{c}
          m_1 \\ n_1 \end{array}\right)
\left(\begin{array}{c}
          m_2 \\ n_2 \end{array}\right)
\left(\begin{array}{c}
          m_3 \\ n_3 \end{array}\right)\0\\\noal
&&\bigg[ \left(\begin{array}{c}
          4d-m_1-2m_2-3m_3-3 \\ 4f-n_1-2n_2-3n_3-1 \end{array}\right)
fg N^f_{n_1+1,n_2,n_3} N^g_{n_1',n_2'+1,n_3'}     \0\\\noal
&&  - \left(\begin{array}{c}
          4d-m_1-2m_2-3m_3-3 \\ 4f-n_1-2n_2-3n_3-1 \end{array}\right)
fg N^f_{n_1,n_2,n_3} N^g_{n_1'+1,n_2'+1,n_3'}     \0\\\noal
&&  - \left(\begin{array}{c}
          4d-m_1-2m_2-3m_3-3 \\ 4f-n_1-2n_2-3n_3-1 \end{array}\right)
f N^f_{n_1+1,n_2,n_3} N^g_{n_1',n_2',n_3'}     \0\\\noal
&&  + \left(\begin{array}{c}
          4d-m_1-2m_2-3m_3-3 \\ 4f-n_1-2n_2-3n_3-1 \end{array}\right)
fN^f_{n_1,n_2,n_3} N^g_{n_1'+1,n_2',n_3'}     \0\\\noal
&&   - \left(\begin{array}{c}
          4d-m_1-2m_2-3m_3-3 \\ 4f-n_1-2n_2-3n_3-1 \end{array}\right)
fN^f_{n_1+2,n_2,n_3} N^g_{n_1'+1,n_2',n_3'}     \0\\\noal
&&\  - \left(\begin{array}{c}
          4d-m_1-2m_2-3m_3-3 \\ 4f-n_1-2n_2-3n_3\end{array}\right)
f^2 N^f_{n_1+1,n_2,n_3} N^g_{n_1',n_2'+1,n_3'}     \0\\\noal
&&  +\left(\begin{array}{c}
          4d-m_1-2m_2-3m_3-3 \\ 4f-n_1-2n_2-3n_3-1 \end{array}\right)
f N^f_{n_1+1,n_2,n_3} N^g_{n_1'+2,n_2',n_3'}\0\\\noal
&&  +\left(\begin{array}{c}
          4d-m_1-2m_2-3m_3-3 \\ 4f-n_1-2n_2-3n_3 \end{array}\right)
f^2 N^f_{n_1,n_2,n_3} N^g_{n_1'+1,n_2'+1,n_3'}    \bigg],\\\noal\noal\noal
&& N^d_{m_1,m_2,m_3+1}+N^d_{m_1+2,m_2,m_3+1}\0\\ \noal
&&=\sum_{\stackrel{f+g=d, n_1+n_1'=m_1} {n_2+n_2'=m_2,
n_3+n_3'=m_3}}\left(\begin{array}{c}
          m_1 \\ n_1 \end{array}\right)
\left(\begin{array}{c}
          m_2 \\ n_2 \end{array}\right)
\left(\begin{array}{c}
          m_3 \\ n_3 \end{array}\right)\0\\\noal
&&\bigg[ \left(\begin{array}{c}
          4d-m_1-2m_2-3m_3-4 \\ 4f-n_1-2n_2-3n_3-2 \end{array}\right)
N^f_{n_1+1,n_2,n_3} N^g_{n_1'+1,n_2',n_3'}     \0\\\noal
&&  - \left(\begin{array}{c}
          4d-m_1-2m_2-3m_3-4 \\ 4f-n_1-2n_2-3n_3-1 \end{array}\right)
fN^f_{n_1+2,n_2,n_3} N^g_{n_1',n_2'+1,n_3'}     \0\\\noal
&&  - \left(\begin{array}{c}
          4d-m_1-2m_2-3m_3-4\\ 4f-n_1-2n_2-3n_3-2 \end{array}\right)
N^f_{n_1,n_2,n_3} N^g_{n_1'+2,n_2',n_3'}     \0\\\noal
&&  - \left(\begin{array}{c}
          4d-m_1-2m_2-3m_3-4 \\ 4f-n_1-2n_2-3n_3-2 \end{array}\right)
N^f_{n_1+2,n_2,n_3} N^g_{n_1'+2,n_2',n_3'}     \0\\\noal
&&   +2\left(\begin{array}{c}
          4d-m_1-2m_2-3m_3-4 \\ 4f-n_1-2n_2-3n_3-1 \end{array}\right)
fN^f_{n_1+1,n_2,n_3} N^g_{n_1'+1,n_2'+1,n_3'}     \0\\\noal
&&\  - \left(\begin{array}{c}
          4d-m_1-2m_2-3m_3-4 \\ 4f-n_1-2n_2-3n_3-2\end{array}\right)
N^f_{n_1+1,n_2,n_3} N^g_{n_1'+3,n_2',n_3'}     \0\\\noal
&&  -\left(\begin{array}{c}
          4d-m_1-2m_2-3m_3-4 \\ 4f-n_1-2n_2-3n_3-1 \end{array}\right)
f N^f_{n_1,n_2,n_3} N^g_{n_1'+2,n_2'+1,n_3'}    \bigg],\\\noal\noal\noal
&& N^d_{m_1+1,m_2+1,m_3+1}\0\\  \noal
&&=\sum_{\stackrel{f+g=d, n_1+n_1'=m_1} {n_2+n_2'=m_2,
n_3+n_3'=m_3}}\left(\begin{array}{c}
          m_1 \\ n_1 \end{array}\right)
\left(\begin{array}{c}
          m_2 \\ n_2 \end{array}\right)
\left(\begin{array}{c}
          m_3 \\ n_3 \end{array}\right)\0\\\noal
&&\bigg[ -\left(\begin{array}{c}
          4d-m_1-2m_2-3m_3-5 \\ 4f-n_1-2n_2-3n_3-2 \end{array}\right)
fN^f_{n_1+1,n_2+1,n_3} N^g_{n_1',n_2'+1,n_3'}     \0\\\noal
&&  +\left(\begin{array}{c}
          4d-m_1-2m_2-3m_3-5 \\ 4f-n_1-2n_2-3n_3-2 \end{array}\right)
N^f_{n_1+1,n_2,n_3} N^g_{n_1',n_2'+1,n_3'}     \0\\\noal
&&  +\left(\begin{array}{c}
          4d-m_1-2m_2-3m_3-5 \\ 4f-n_1-2n_2-3n_3-2 \end{array}\right)
f N^f_{n_1,n_2+1,n_3} N^g_{n_1'+1,n_2'+1,n_3'}     \0\\\noal
&&  -\left(\begin{array}{c}
          4d-m_1-2m_2-3m_3-5 \\ 4f-n_1-2n_2-3n_3-2 \end{array}\right)
N^f_{n_1,n_2,n_3} N^g_{n_1'+1,n_2'+1,n_3'}     \0\\\noal
&&   +\left(\begin{array}{c}
          4d-m_1-2m_2-3m_3-5 \\ 4f-n_1-2n_2-3n_3-2 \end{array}\right)
N^f_{n_1+2,n_2,n_3} N^g_{n_1'+1,n_2'+1,n_3'}     \0\\\noal
&&\  +\left(\begin{array}{c}
          4d-m_1-2m_2-3m_3-5 \\ 4f-n_1-2n_2-3n_3-1\end{array}\right)
fN^f_{n_1+1,n_2,n_3} N^g_{n_1',n_2'+2,n_3'}     \0\\\noal
&&  -\left(\begin{array}{c}
          4d-m_1-2m_2-3m_3-5 \\ 4f-n_1-2n_2-3n_3-2 \end{array}\right)
N^f_{n_1+1,n_2,n_3} N^g_{n_1'+2,n_2'+1,n_3'}\0\\\noal
&&  -\left(\begin{array}{c}
          4d-m_1-2m_2-3m_3-5 \\ 4f-n_1-2n_2-3n_3-1 \end{array}\right)
fN^f_{n_1,n_2,n_3} N^g_{n_1'+1,n_2'+2,n_3'}    \bigg],\\\noal\noal\noal
&& N^d_{m_1+1,m_2,m_3+2}\0\\
&&=\sum_{\stackrel{f+g=d, n_1+n_1'=m_1} {n_2+n_2'=m_2,
n_3+n_3'=m_3}}\left(\begin{array}{c}
          m_1 \\ n_1 \end{array}\right)
\left(\begin{array}{c}
          m_2 \\ n_2 \end{array}\right)
\left(\begin{array}{c}
          m_3 \\ n_3 \end{array}\right)\0\\\noal
&&\bigg[ -\left(\begin{array}{c}
          4d-m_1-2m_2-3m_3-6 \\ 4f-n_1-2n_2-3n_3-3 \end{array}\right)
fN^f_{n_1+1,n_2,n_3+1} N^g_{n_1',n_2'+1,n_3'}     \0\\\noal
&&  + \left(\begin{array}{c}
          4d-m_1-2m_2-3m_3-6 \\ 4f-n_1-2n_2-3n_3-2 \end{array}\right)
N^f_{n_1+1,n_2,n_3} N^g_{n_1',n_2',n_3'+1}     \0\\\noal
&&  + \left(\begin{array}{c}
          4d-m_1-2m_2-3m_3-6 \\ 4f-n_1-2n_2-3n_3-3 \end{array}\right)
f N^f_{n_1,n_2,n_3+1} N^g_{n_1'+1,n_2'+1,n_3'}    \0\\\noal
&&  -\left(\begin{array}{c}
          4d-m_1-2m_2-3m_3-6 \\ 4f-n_1-2n_2-3n_3-2 \end{array}\right)
N^f_{n_1,n_2,n_3} N^g_{n_1'+1,n_2',n_3'+1}     \0\\\noal
&&   +\left(\begin{array}{c}
          4d-m_1-2m_2-3m_3-6 \\ 4f-n_1-2n_2-3n_3-2 \end{array}\right)
N^f_{n_1+1,n_2,n_3} N^g_{n_1'+1,n_2',n_3'+1}     \0\\\noal
&&\  +\left(\begin{array}{c}
          4d-m_1-2m_2-3m_3-6 \\ 4f-n_1-2n_2-3n_3-1\end{array}\right)
fN^f_{n_1+1,n_2,n_3} N^g_{n_1',n_2'+1,n_3'+1}     \0\\\noal
&&  -\left(\begin{array}{c}
          4d-m_1-2m_2-3m_3-6 \\ 4f-n_1-2n_2-3n_3-2 \end{array}\right)
N^f_{n_1+1,n_2,n_3} N^g_{n_1'+2,n_2',n_3'+1}\0\\\noal
&&  -\left(\begin{array}{c}
          4d-m_1-2m_2-3m_3-6 \\ 4f-n_1-2n_2-3n_3-1\end{array}\right)
fN^f_{n_1,n_2,n_3} N^g_{n_1'+1,n_2'+1,n_3'+1}    \bigg],\\\noal\noal\noal
&& N^d_{m_1,m_2+3,m_3}-N^d_{m_1,m_2+1,m_3+1}\0\\  \noal
&&=\sum_{\stackrel{f+g=d, n_1+n_1'=m_1} {n_2+n_2'=m_2,
n_3+n_3'=m_3}}\left(\begin{array}{c}
          m_1 \\ n_1 \end{array}\right)
\left(\begin{array}{c}
          m_2 \\ n_2 \end{array}\right)
\left(\begin{array}{c}
          m_3 \\ n_3 \end{array}\right)\0\\\noal
&&\bigg[ \left(\begin{array}{c}
          4d-m_1-2m_2-3m_3-5 \\ 4f-n_1-2n_2-3n_3-2 \end{array}\right)
fN^f_{n_1,n_2+1,n_3} N^g_{n_1',n_2'+1,n_3'}     \0\\\noal
&&  +\left(\begin{array}{c}
          4d-m_1-2m_2-3m_3-5 \\ 4f-n_1-2n_2-3n_3-2 \end{array}\right)
fN^f_{n_1+1,n_2+1,n_3} N^g_{n_1'+1,n_2'+1,n_3'}     \0\\\noal
&&  +\left(\begin{array}{c}
          4d-m_1-2m_2-3m_3-5 \\ 4f-n_1-2n_2-3n_3-1 \end{array}\right)
f^2 N^f_{n_1,n_2+1,n_3} N^g_{n_1',n_2'+2,n_3'}     \0\\\noal
&&  -\left(\begin{array}{c}
          4d-m_1-2m_2-3m_3-5 \\ 4f-n_1-2n_2-3n_3-1 \end{array}\right)
fN^f_{n_1,n_2,n_3} N^g_{n_1',n_2'+2,n_3'}     \0\\\noal
&&   - \left(\begin{array}{c}
          4d-m_1-2m_2-3m_3-5 \\ 4f-n_1-2n_2-3n_3-1 \end{array}\right)
fN^f_{n_1+1,n_2,n_3} N^g_{n_1'+1,n_2'+2,n_3'}     \0\\\noal
&&\  - \left(\begin{array}{c}
          4d-m_1-2m_2-3m_3-5 \\ 4f-n_1-2n_2-3n_3\end{array}\right)
f^2 N^f_{n_1,n_2,n_3} N^g_{n_1',n_2'+2,n_3'}         \bigg],\\\noal\noal\noal
&& N^d_{m_1,m_2+2,m_3}-N^d_{m_1+1,m_2,m_3+1}\0\\  \noal
&&=\sum_{\stackrel{f+g=d, n_1+n_1'=m_1} {n_2+n_2'=m_2,
n_3+n_3'=m_3}}\left(\begin{array}{c}
          m_1 \\ n_1 \end{array}\right)
\left(\begin{array}{c}
          m_2 \\ n_2 \end{array}\right)
\left(\begin{array}{c}
          m_3 \\ n_3 \end{array}\right)\0\\\noal
&&\bigg[ -\left(\begin{array}{c}
          4d-m_1-2m_2-3m_3-4 \\ 4f-n_1-2n_2-3n_3-2 \end{array}\right)
fg N^f_{n_1,n_2+1,n_3} N^g_{n_1'+1,n_2'+1,n_3'}     \0\\\noal
&&  +\left(\begin{array}{c}
          4d-m_1-2m_2-3m_3-4 \\ 4f-n_1-2n_2-3n_3-1 \end{array}\right)
fg N^f_{n_1+1,n_2,n_3} N^g_{n_1',n_2'+2,n_3'}     \0\\\noal
&&  +\left(\begin{array}{c}
          4d-m_1-2m_2-3m_3-4 \\ 4f-n_1-2n_2-3n_3-2 \end{array}\right)
f N^f_{n_1,n_2+1,n_3} N^g_{n_1'+1,n_2',n_3'}     \0\\\noal
&&  -\left(\begin{array}{c}
          4d-m_1-2m_2-3m_3-4 \\ 4f-n_1-2n_2-3n_3-1 \end{array}\right)
fN^f_{n_1+1,n_2,n_3} N^g_{n_1',n_2'+1,n_3'}     \0\\\noal
&&   +\left(\begin{array}{c}
          4d-m_1-2m_2-3m_3-4 \\ 4f-n_1-2n_2-3n_3-2 \end{array}\right)
fN^f_{n_1+1,n_2+1,n_3} N^g_{n_1'+2,n_2',n_3'}     \0\\\noal
&&\  +\left(\begin{array}{c}
          4d-m_1-2m_2-3m_3-4 \\ 4f-n_1-2n_2-3n_3-1\end{array}\right)
f^2 N^f_{n_1,n_2+1,n_3} N^g_{n_1'+1,n_2'+1,n_3'}     \0\\\noal
&&  -\left(\begin{array}{c}
          4d-m_1-2m_2-3m_3-4 \\ 4f-n_1-2n_2-3n_3-1 \end{array}\right)
f N^f_{n_1+2,n_2,n_3} N^g_{n_1'+1,n_2'+1,n_3'}\0\\\noal
&&  -\left(\begin{array}{c}
          4d-m_1-2m_2-3m_3-4 \\ 4f-n_1-2n_2-3n_3 \end{array}\right)
f^2 N^f_{n_1+1,n_2,n_3} N^g_{n_1',n_2'+2,n_3'}    \bigg],\\\noal \noal\noal
&& N^d_{m_1,m_2+1,m_3+2}\0\\  \noal
&&=\sum_{\stackrel{f+g=d, n_1+n_1'=m_1} {n_2+n_2'=m_2,
n_3+n_3'=m_3}}\left(\begin{array}{c}
          m_1 \\ n_1 \end{array}\right)
\left(\begin{array}{c}
          m_2 \\ n_2 \end{array}\right)
\left(\begin{array}{c}
          m_3 \\ n_3 \end{array}\right)\0\\\noal
&&\bigg[ \left(\begin{array}{c}
          4d-m_1-2m_2-3m_3-7 \\ 4f-n_1-2n_2-3n_3-3 \end{array}\right)
fN^f_{n_1,n_2,n_3+1} N^g_{n_1',n_2'+2,n_3'}     \0\\\noal
&&  - \left(\begin{array}{c}
          4d-m_1-2m_2-3m_3-7 \\ 4f-n_1-2n_2-3n_3-2 \end{array}\right)
fN^f_{n_1,n_2+1,n_3} N^g_{n_1',n_2'+1,n_3'+1}     \0\\\noal
&&  +\left(\begin{array}{c}
          4d-m_1-2m_2-3m_3-7 \\ 4f-n_1-2n_2-3n_3-3 \end{array}\right)
f N^f_{n_1+1,n_2,n_3+1} N^g_{n_1'+1,n_2'+2,n_3'}     \0\\\noal
&&  -\left(\begin{array}{c}
          4d-m_1-2m_2-3m_3-7 \\ 4f-n_1-2n_2-3n_3-2 \end{array}\right)
fN^f_{n_1+1,n_2+1,n_3} N^g_{n_1'+1,n_2'+1,n_3'+1}     \0\\\noal
&&   +\left(\begin{array}{c}
          4d-m_1-2m_2-3m_3-7 \\ 4f-n_1-2n_2-3n_3-2 \end{array}\right)
f^2N^f_{n_1,n_2,n_3+1} N^g_{n_1',n_2'+3,n_3'}     \0\\\noal
&&\  - \left(\begin{array}{c}
          4d-m_1-2m_2-3m_3-7 \\ 4f-n_1-2n_2-3n_3-1\end{array}\right)
f^2 N^f_{n_1,n_2+1,n_3} N^g_{n_1',n_2'+2,n_3'+1}         \bigg].
\bj
\normalsize
In there recursion relations, d, f, and g are all greater or equal to one. So
when d equals one, r,h,s of these equations vanish since g+f $\geq$ 2. Then
we have a set of linear relations for $N^1_*$'s. We can use these linear
relations to determine all the the 
$<{\cal O}_{W_4}{\cal O}_{W_4}>$=1 for $CP^3$, $<{\cal O}_{W_5}{\cal
O}_{W_5}>$=1 for $CP^4$ and $<{\cal O}_{W_5}{\cal O}_{W_6}>$=1 for $Gr(2,4)$.
Then we put
these degree 1 correlation functions to the r,h,s of (4.35),(4.36),(4.37)
and obtain linear
relations for $N^2_*$'s. This time, these linear relations thoroughly determine
them. For higher degree, the process is the same as d=2 case. We observe that
recursion relations we have witten down are suffcient for determination.
We checked the compatiable condition in the case of d $\leq$ 4. It seems that
the over determined system of WDVV equation work well in all degrees of maps in
the case of $CP^3, CP^4$ and $Gr(2,4)$. The intersection numbers of moduli
space of d $\leq$ 4 are given in the following tables.
\footnotesize
\begin{table}[h]
\caption{\bf D=1 $CP^3$}
  \begin{center}
    \begin{tabular}{|l|l|l|}
     \hline
$N_{0}=2\;\;\;$ & $N_{1}=1\;\;\;$ & $N_{2}=1\;\;\;$ \\ \hline
\end{tabular}
      \end{center}
     \end{table}
\vskip-1.5cm
\begin{table}
\caption{D=2 $CP^3$}
  \begin{center}
    \begin{tabular}{|l|l|l|l|l|l|}
     \hline
$N_{0}=92\;\;\;$ & $N_{1}=18\;\;\;$ & $N_{2}=4\;\;\;$ & $N_{3}=1\;\;\;$ &
$N_{4}=0\;\;\;$ \\ \hline
\end{tabular}
      \end{center}
     \end{table}
\vskip-1.5cm
\begin{table}
\caption{D=3 $CP^3$}
  \begin{center}
    \begin{tabular}{|l|l|l|l|l|l|l|l|}
     \hline
$N_{0}=80160$ & $N_{1}=9864$ & $N_{2}=1312$ & $N_{3}=190$ & $N_{4}=30$ &
$N_{5}=5$ & $N_{6}=1$ \\ \hline
\end{tabular}
      \end{center}
     \end{table}
\vskip-1.5cm
\begin{table}
\caption{D=4 $CP^3$}
  \begin{center}
    \begin{tabular}{|l|l|l|l|l|}
     \hline
$N_{0}=383306880\;$ & $N_{1}=34382544\;$ & $N_{2}=3259680\;$ & $N_{3}=327888\;$
& $N_{4}=35104\;$ \\ \hline
$N_{5}=4000\;$ & $N_{6}=480\;$ & $N_{7}=58\;$ & $N_{8}=4\;$ & \\ \hline
\end{tabular}
      \end{center}
     \end{table}
\vskip-1.5cm
\begin{table}
\caption{D=5 $CP^3$}
  \begin{center}
    \begin{tabular}{|l|l|l|l|}
     \hline
$N_{0}=6089786376960$ & $N_{1}=429750191232$ & $N_{2}=31658432256$ &
$N_{3}=2440235712$  \\ \hline
$N_{4}=197240400$ & $N_{5}=16744080$ & $N_{6}=1492616$ & $N_{7}=139098$ \\
\hline
 $N_{8}=13354$ & $N_{9}=1265$ & $N_{10}=105$ & \\ \hline
\end{tabular}
      \end{center}
     \end{table}
\vskip-1.5cm
\begin{table}
\caption{D=6 $CP^3$}
  \begin{center}
    \begin{tabular}{|l|l|l|l|}
     \hline
$N_{0}=244274488980962304\;$ & $N_{1}=14207926965714432\;$ &
$N_{2}=855909223176192\;$ \\ \hline
$N_{3}=53486265350784\;$ & $N_{4}=3472451647488\;$ & $N_{5}=234526910784\;$ \\
\hline
$N_{6}=16492503552\;$ & $N_{7}=1207260512\;$ & $N_{8}=91797312\;$ \\ \hline
$N_{9}=7200416\;$ & $N_{10}=573312\;$  & $N_{11}=44416\;$ \\ \hline
$N_{12}=2576$ & &\\ \hline
\end{tabular}
      \end{center}
     \end{table}
   \clearpage
\begin{table}
\caption{D=1 $CP^4$}
  \begin{center}
    \begin{tabular}{|l|l|l|l|l|l|l|}
     \hline
$N_{00}=5\;\;$ & $N_{10}=3\;\;$ & $N_{20}=2\;\;$ & $N_{30}=1\;\;$ &
$N_{01}=1\;\;$ & $N_{11}=1\;\;$ & $N_{02}=1\;\;$ \\ \hline
\end{tabular}~
      \end{center}
     \end{table}

\begin{table}
\caption{D=2 $CP^4$}
  \begin{center}
    \begin{tabular}{|l|l|l|l|l|l|}
     \hline
$N_{00}=6620\;$ & $N_{10}=1734\;$ & $N_{20}=473\;$ & $N_{30}=132\;$ &
$N_{40}=36\;$ & $N_{50}=10\;$ \\ \hline
$N_{01}=219$ & $N_{11}=67$ & $N_{21}=21$ & $N_{31}=6$ & $N_{41}=2$ & \\ \hline
$N_{02}=11$ & $N_{12}=4$ & $N_{22}=1$ & & &\\ \hline
$N_{03}=1$ & $N_{13}=0$ & & & &\\ \hline
\end{tabular}
      \end{center}
     \end{table}
\vskip-1cm
\begin{table}
\caption{D=3 $CP^4$}
  \begin{center}
    \begin{tabular}{|l|l|l|l|l|}
     \hline
$N_{00}=213709980$ & $N_{01}=2770596$ & $N_{02}=45954$ & $N_{03}=1011$ &
$N_{04}=30$  \\ \hline
$N_{05}=0$ & $N_{10}=35806494$ & $N_{11}=511012$ & $N_{12}=9386$ & $N_{13}=225$
  \\ \hline
$N_{14}=5$ & $N_{20}=6165822$ & $N_{21}=96548$ & $N_{22}=1931$ & $N_{23}=45$
\\ \hline
$N_{24}=1$ & $N_{30}=1085892$ & $N_{31}=18469$ & $N_{32}=385$ & $N_{33}=9$   \\
\hline
$N_{40}=194024$ & $N_{41}=3512$ & $N_{42}=76$ & & \\ \hline
$N_{50}=34780$ & $N_{51}=664$ & $N_{52}=16$ & & \\ \hline
$N_{60}=6216$ & $N_{61}=128$ & & & \\ \hline
$N_{70}=1108$ & & &  &\\ \hline
$N_{80}=188$ & &  & &\\ \hline
\end{tabular}
      \end{center}
     \end{table}
\vskip -1cm
\begin{table}
\caption{D=4 $CP^4$}
  \begin{center}
    \begin{tabular}{|l|l|l|l|}
     \hline
$N_{00}=47723447905060$ & $N_{01}=327439797532$ & $N_{02}=2679044142$ &
$N_{03}=26578256$ \\ \hline
$N_{04}=324764$ & $N_{05}=4830$ & $N_{06}=61$ & $N_{07}=1$  \\ \hline
$N_{10}=5876564125104$ & $N_{11}=43242657488$ & $N_{12}=380720598$ &
$N_{13}=4063860$ \\ \hline
$N_{14}=52507$ & $N_{15}=732$ & $N_{16}=9$ & \\ \hline
$N_{20}=738764469204$ & $N_{21}=5823161346$ & $N_{22}=54948346$ &
$N_{23}=622980$  \\ \hline
$N_{24}=8133$ & $N_{25}=107$ & &\\ \hline
$N_{30}=94605276228$ & $N_{31}=796460052$ & $N_{32}=7990720$ & $N_{33}=94104$
\\ \hline
$N_{34}=1218$ & $N_{35}=14$ & &\\ \hline
$N_{40}=12302188692$ & $N_{41}=110031632$ & $N_{42}=1159218$ & $N_{43}=13962$
\\ \hline
$N_{44}=178$ & & &\\ \hline
$N_{50}=1617593360$ & $N_{51}=15251816$ & $N_{52}=166936$ & $N_{53}=2056$ \\
\hline
$N_{60}=213984472$ & $N_{61}=2110864$ & $N_{62}=23968$ & $N_{63}=320$ \\ \hline
$N_{70}=28346212$ & $N_{71}=291632$ & $N_{72}=3516$ & \\ \hline
$N_{80}=3748804$ & $N_{81}=40492$ & & \\ \hline
$N_{90}=343260$ & $N_{91}=5552$ & & \\ \hline
$N_{100}=63740$ & & &\\ \hline
       \end{tabular}
      \end{center}
     \end{table}


\begin{table}
\caption{D=1 Gr(2,4)}
  \begin{center}
    \begin{tabular}{|l|l|l|l|l|l|}
     \hline
$N_{000}=0\;\;\;\;\;\;\;\;$ & $N_{100}=0\;\;\;\;\;\;\;\;$ &
$N_{200}=1\;\;\;\;\;\;\;\;$ & $N_{300}=1\;\;\;\;\;\;\;\;$ &
$N_{400}=0\;\;\;\;\;\;\;\;$ & $N_{500}=0\;\;\;\;\;\;\;\;$ \\ \hline
$N_{001}=0$ & $N_{101}=1$ & $N_{201}=0$ & & & \\ \hline
$N_{010}=0$ & $N_{110}=1$ & $N_{210}=1$ & $N_{310}=0$ & & \\ \hline
$N_{011}=1$ & & & & & \\ \hline
$N_{020}=1$ & $N_{120}=1$ & & & & \\ \hline
\end{tabular}
      \end{center}
     \end{table}

\vskip-30cm

\begin{table}
\caption{D=2 Gr(2,4)}
  \begin{center}
    \begin{tabular}{|l|l|l|l|l|l|l|l|}
     \hline
$N_{000}=2$ & $N_{100}=6$ & $N_{200}=18$ & $N_{300}=34$ & $N_{400}=42$ &
$N_{500}=42$ & $N_{600}=34$ & $N_{700}=18$ \\ \hline
$N_{800}=6$ & $N_{900}=2$ & & & & & &\\ \hline
$N_{001}=1$ & $N_{101}=3$ & $N_{201}=5$ & $N_{301}=5$ & $N_{401}=5$ &
$N_{501}=3$ & $N_{601}=1$ &\\ \hline
$N_{002}=1$ & $N_{102}=1$ & $N_{202}=1$ & $N_{302}=1$ & & & &\\ \hline
$N_{003}=1$ & & & & & & & \\ \hline
$N_{010}=3$ & $N_{110}=9$ & $N_{210}=17$ & $N_{310}=21$ & $N_{410}=21$ &
$N_{510}=17$ & $N_{610}=9$ & $N_{710}=3$\\ \hline
$N_{011}=2$ & $N_{111}=3$ & $N_{211}=3$ & $N_{311}=3$ & $N_{411}=2$ & & & \\
\hline
$N_{012}=1$ & $N_{112}=1$ & & & & & & \\ \hline
$N_{020}=5$ & $N_{120}=9$ & $N_{220}=11$ & $N_{320}=11$ & $N_{420}=9$ &
$N_{520}=5$ & & \\ \hline
$N_{021}=2$ & $N_{121}=2$ & $N_{221}=2$ & & & & & \\ \hline
$N_{030}=5$ & $N_{130}=6$ & $N_{230}=6$ & $N_{330}=5$ & & & & \\ \hline
$N_{031}=1$ & & & & & & & \\ \hline
$N_{040}=3$ & $N_{140}=3$ & & & & & & \\ \hline
       \end{tabular}
      \end{center}
     \end{table}

\begin{table}
\caption{D=3 Gr(2,4)}
  \begin{center}
    \begin{tabular}{|l|l|l|l|l|}
     \hline
\vspace{-0.5mm}
$N_{000}=504\;\;\;\;\;\;\;\;\;\;\;\;$ & $N_{001}=100\;\;\;\;\;\;\;\;\;\;\;\;$ &
$N_{002}=25\;\;\;\;\;\;\;\;\;\;\;\;$ & $N_{003}=6\;\;\;\;\;\;\;\;\;\;\;\;$ &
$N_{004}=2\;\;\;\;\;\;\;\;\;\;\;\;$ \\ \hline
\vspace{-0.5mm}
$N_{100}=1824$ & $N_{101}=307$ & $N_{102}=55$ & $N_{103}=9$ & $N_{104}=2$  \\
\hline
\vspace{-0.5mm}
$N_{200}=5159$ & $N_{201}=676$ & $N_{202}=83$ & $N_{203}=10$ & \\ \hline
\vspace{-0.5mm}
$N_{300}=11319$ & $N_{301}=1109$ & $N_{302}=101$ & $N_{303}=9$ & \\ \hline
\vspace{-0.5mm}
$N_{400}=19512$ & $N_{401}=1460$ & $N_{402}=101$ & $N_{403}=6$ & \\ \hline
\vspace{-0.5mm}
$N_{500}=27472$ & $N_{501}=1605$ & $N_{502}=83$ & &\\ \hline
\vspace{-0.5mm}
$N_{600}=32517$ & $N_{601}=1460$ & $N_{602}=55$ & & \\ \hline
\vspace{-0.5mm}
$N_{700}=32517$ & $N_{701}=1109$ & $N_{702}=25$ & & \\ \hline
\vspace{-0.5mm}
$N_{800}=27472$ & $N_{801}=676$ & & & \\ \hline
\vspace{-0.5mm}
$N_{900}=19512$ & $N_{901}=307$ & & & \\ \hline
\vspace{-0.5mm}
$N_{1000}=11319$ & $N_{1001}=100$ & & & \\ \hline
\vspace{-0.5mm}
$N_{1100}=5159$ & & & & \\ \hline
\vspace{-0.5mm}
$N_{1200}=1824$ & & & &\\ \hline
\vspace{-0.5mm}
$N_{1300}=504$ & & & &\\ \hline
\vspace{-0.5mm}
$N_{010}=538$ & $N_{011}=109$ & $N_{012}=23$ & $N_{013}=5$ &  \\ \hline
\vspace{-0.5mm}
$N_{110}=1603$ & $N_{111}=246$ & $N_{112}=35$ & $N_{113}=5$ &  \\ \hline
\vspace{-0.5mm}
$N_{210}=3607$ & $N_{211}=403$ & $N_{212}=42$ & $N_{213}=5$ &  \\ \hline
\vspace{-0.5mm}
$N_{310}=6278$ & $N_{311}=528$ & $N_{312}=42$ & & \\ \hline
\vspace{-0.5mm}
$N_{410}=8864$ & $N_{411}=579$ & $N_{412}=35$ & & \\ \hline
\vspace{-0.5mm}
$N_{510}=10499$ & $N_{511}=528$ & $N_{512}=23$ & & \\ \hline
\vspace{-0.5mm}
$N_{610}=10499$ & $N_{611}=403$ & & &\\ \hline
\vspace{-0.5mm}
$N_{710}=8864$ & $N_{711}=246$ & & & \\ \hline
\vspace{-0.5mm}
$N_{810}=6278$ & $N_{811}=109$ & & & \\ \hline
\vspace{-0.5mm}
$N_{910}=3609$ & & & & \\ \hline
\vspace{-0.5mm}
$N_{1010}=1603$ & & & & \\ \hline
\vspace{-0.5mm}
$N_{1110}=538$ & & & & \\ \hline
\vspace{-0.5mm}
$N_{020}=523$ & $N_{021}=94$ & $N_{022}=16$ & $N_{023}=2$ &\\ \hline
\vspace{-0.5mm}
$N_{120}=1203$ & $N_{121}=153$ & $N_{122}=18$ & & \\ \hline
\vspace{-0.5mm}
$N_{220}=2100$ & $N_{221}=198$ & $N_{222}=18$ & & \\ \hline
\vspace{-0.5mm}
$N_{320}=2960$ & $N_{321}=216$ & $N_{322}=16$ & & \\ \hline
\vspace{-0.5mm}
$N_{420}=3501$ & $N_{421}=198$ & & & \\ \hline
\vspace{-0.5mm}
$N_{520}=3501$ & $N_{521}=153$ & & & \\ \hline
\vspace{-0.5mm}
$N_{620}=2960$ & $N_{621}=94$ & & & \\ \hline
\vspace{-0.5mm}
$N_{720}=2100$ & & & & \\ \hline
\vspace{-0.5mm}
$N_{820}=1203$ & & & & \\ \hline
\vspace{-0.5mm}
$N_{920}=523$ & & & & \\ \hline
\vspace{-0.5mm}
$N_{030}=420$ & $N_{031}=61$ & $N_{032}=7$ & &\\ \hline
\vspace{-0.5mm}
$N_{130}=729$ & $N_{131}=76$ & $N_{132}=7$ & &\\ \hline
\vspace{-0.5mm}
$N_{230}=1019$ & $N_{231}=82$ & & &\\ \hline
\vspace{-0.5mm}
$N_{330}=1200$ & $N_{331}=76$ & & &\\ \hline
\vspace{-0.5mm}
$N_{430}=1200$ & $N_{431}=61$ & & &\\ \hline
\vspace{-0.5mm}
$N_{530}=1019$ & & & &\\ \hline
\vspace{-0.5mm}
$N_{630}=729$ & & & &\\ \hline
\vspace{-0.5mm}
$N_{730}=420$ & & & &\\ \hline
\vspace{-0.5mm}
$N_{040}=262$ & $N_{041}=28$ & & &\\ \hline
\vspace{-0.5mm}
$N_{140}=358$ & $N_{141}=30$ & & &\\ \hline
\vspace{-0.5mm}
$N_{240}=418$ & $N_{241}=28$ & & &\\ \hline
\vspace{-0.5mm}
$N_{340}=418$ & & & &\\ \hline
\vspace{-0.5mm}
$N_{440}=358$ & & & &\\ \hline
\vspace{-0.5mm}
$N_{540}=262$ & & & &\\ \hline
\vspace{-0.5mm}
$N_{050}=124$ & $N_{051}=10$ & & &\\ \hline
\vspace{-0.5mm}
$N_{150}=144$ & & & &\\ \hline
\vspace{-0.5mm}
$N_{250}=144$ & & & &\\ \hline
\vspace{-0.5mm}
$N_{350}=124$ & & & &\\ \hline
\vspace{-0.5mm}
$N_{060}=48$ & & & &\\ \hline
\vspace{-0.5mm}
$N_{160}=48$ & & & &\\ \hline
\end{tabular}
      \end{center}
     \end{table}

\begin{table}
\caption{D=4 Gr(2,4)}
  \begin{center}
    \begin{tabular}{|l|l|l|l|l|l|}
     \hline
$N_{000}=1044120$ & $N_{001}=93726$ & $N_{002}=9970$ & $N_{003}=1170$ &
$N_{004}=138$ & $N_{005}=20$ \\ \hline
$N_{100}=3094440$ & $N_{101}=251402$ & $N_{102}=22570$ & $N_{103}=2058$ &
$N_{104}=190$ & $N_{105}=20$ \\ \hline
$N_{200}=8093840$ & $N_{201}=570998$ & $N_{202}=4179$ & $N_{203}=2998$ &
$N_{204}=214$ & $N_{205}=20$ \\ \hline
$N_{300}=18245976$ & $N_{301}=1086890$ & $N_{302}=64434$ & $N_{303}=3690$ &
$N_{304}=214$ &\\ \hline
$N_{400}=35219976$ & $N_{401}=1752446$ & $N_{402}=84818$ & $N_{403}=3942$ &
$N_{404}=190$ &\\ \hline
$N_{500}=58571280$ & $N_{501}=2434530$ & $N_{502}=96894$ & $N_{503}=3690$ &
$N_{504}=138$ &\\ \hline
$N_{600}=84843440$ & $N_{601}=2951174$ & $N_{602}=96894$ & $N_{603}=2998$ & &
\\ \hline
$N_{700}=108066120$ & $N_{701}=3143726$ & $N_{702}=84818$ & $N_{703}=2058$ & &
\\ \hline
$N_{800}=121770480$ & $N_{801}=2951174$ & $N_{802}=64434$ & $N_{803}=1170$ & &
\\ \hline
$N_{900}=121770480$ & $N_{901}=2434530$ & $N_{902}=41794$ & & & \\ \hline
$N_{1000}=108066120$ & $N_{1001}=1752446$ & $N_{1002}=22570$ & & & \\ \hline
$N_{1100}=84843440$ & $N_{1101}=1086890$ & $N_{1102}=9970$ & & & \\ \hline
$N_{1200}=58571280$ & $N_{1201}=570998$ & & & &\\ \hline
$N_{1300}=35219976$ & $N_{1301}=251402$ & & & &\\ \hline
$N_{1400}=18245976$ & $N_{1401}=93726$  & & & & \\ \hline
$N_{1500}=8093840$ & & & & &\\ \hline
$N_{1600}=3094440$ & & & & &\\ \hline
$N_{1700}=1044120$ & & & & &\\ \hline
$N_{010}=692760$ & $N_{011}=63904$ & $N_{012}=6528$ & $N_{013}=675$ &
$N_{014}=74$ & $N_{015}=6$ \\ \hline
$N_{110}=1852184$ & $N_{111}=147070$ & $N_{112}=12060$ & $N_{113}=976$ &
$N_{114}=80$ & \\ \hline
$N_{210}=4249660$ & $N_{211}=281764$ & $N_{212}=18506$ & $N_{213}=1181$ &
$N_{214}=80$ &  \\ \hline
$N_{310}=8297556$ & $N_{311}=454858$ & $N_{312}=24196$ & $N_{313}=1254$ &
$N_{314}=74$ & \\ \hline
$N_{410}=13886500$ & $N_{411}=631136$ & $N_{412}=27526$ & $N_{413}=1181$ & & \\
\hline
$N_{510}=20177804$ & $N_{511}=764000$ & $N_{512}=27526$ & $N_{513}=976$ & & \\
\hline
$N_{610}=25736664$ & $N_{611}=813396$ & $N_{612}=24196$ & $N_{613}=675$ & &\\
\hline
$N_{710}=29015656$ & $N_{711}=764000$ & $N_{712}=18506$ & & & \\ \hline
$N_{810}=29015656$ & $N_{811}=631136$ & $N_{812}=12060$ & & & \\ \hline
$N_{910}=25736664$ & $N_{911}=454858$ & $N_{912}=6528$ & & & \\ \hline
$N_{1010}=20177804$ & $N_{1011}=281764$ & & & & \\ \hline
$N_{1110}=13886500$ & $N_{1111}=147070$ & & & & \\ \hline
$N_{1210}=8297556$ & $N_{1211}=63904$ & & & & \\ \hline
$N_{1310}=4249660$ & & & & & \\ \hline
$N_{1410}=1852184$ & & & & & \\ \hline
$N_{1510}=692760$ & & & & & \\ \hline
$N_{020}=440638$ & $N_{021}=39460$ & $N_{022}=3624$ & $N_{023}=332$ &
$N_{024}=26$ & \\ \hline
$N_{120}=1025894$ & $N_{121}=75712$ & $N_{122}=5512$ & $N_{123}=338$ &
$N_{124}=26$ & \\ \hline
$N_{220}=2019894$ & $N_{221}=121884$ & $N_{222}=7112$ & $N_{223}=408$ & &\\
\hline
$N_{320}=3391958$ & $N_{321}=168332$ & $N_{322}=8032$ & $N_{323}=338$ & & \\
\hline
$N_{420}=4932358$ & $N_{421}=203048$ & $N_{422}=8032$ & $N_{423}=332$ &  &\\
\hline
$N_{520}=6290046$ & $N_{521}=215904$ & $N_{522}=7112$ & & & \\ \hline
$N_{620}=7089646$ & $N_{621}=203048$ & $N_{622}=5512$ & &  &\\ \hline
$N_{720}=7089646$ & $N_{721}=168332$ & $N_{722}=3624$ & & & \\ \hline
$N_{820}=6290046$ & $N_{821}=121884$ & & & &\\ \hline
$N_{920}=4932358$ & $N_{921}=75712$ & & &  &\\ \hline
$N_{1020}=3391958$ & $N_{1021}=39460$ & & & & \\ \hline
$N_{1120}=2019894$ & & & & &\\ \hline
$N_{1220}=1025894$ & & & & & \\ \hline
$N_{1320}=440638$ & & & & & \\ \hline
       \end{tabular}
      \end{center}
     \end{table}

\begin{table}
\caption{D=4 Gr(2,4)}
  \begin{center}
    \begin{tabular}{|l|l|l|l|l|l|}
     \hline
$N_{030}=256946$ & $N_{031}=21072$ & $N_{032}=1695$ & $N_{033}=121$ & \ \ \ \ \
\ \ \ \ \ \ \ \ \ & \ \ \ \ \ \ \ \ \ \ \ \ \ \  \\ \hline
$N_{130}=508026$ & $N_{131}=33665$ & $N_{132}=2131$ & $N_{133}=126$ & &\\
\hline
$N_{230}=852818$ & $N_{231}=46042$ & $N_{232}=2379$ & $N_{233}=121$ & &\\
\hline
$N_{330}=1237234$ & $N_{331}=55181$ & $N_{332}=2379$ & & &\\ \hline
$N_{430}=1574370$ & $N_{431}=58548$ & $N_{432}=2131$ & & &\\ \hline
$N_{530}=1772374$ & $N_{531}=55181$ & $N_{532}=1695$ & & &\\ \hline
$N_{630}=1772374$ & $N_{631}=46042$ & & & &\\ \hline
$N_{730}=1574370$ & $N_{731}=33665$ & & & &\\ \hline
$N_{830}=1237234$ & $N_{831}=21072$ & & & &\\ \hline
$N_{930}=852818$ & & & & &\\ \hline
$N_{1030}=508026$ & & & & &\\ \hline
$N_{1130}=256946$ & & & & &\\ \hline
$N_{040}=131874$ & $N_{041}=9540$ & $N_{042}=626$ & $N_{043}=36$ & &\\ \hline
$N_{140}=220250$ & $N_{141}=12808$ & $N_{142}=690$ & & &\\ \hline
$N_{240}=317466$ & $N_{241}=15196$ & $N_{242}=690$ & & &\\ \hline
$N_{340}=402090$ & $N_{341}=16072$ & $N_{342}=626$ & & &\\ \hline
$N_{440}=451610$ & $N_{441}=15196$ & & & &\\ \hline
$N_{540}=451610$ & $N_{541}=12808$ & & & &\\ \hline
$N_{640}=402090$ & $N_{641}=9540$ & & & &\\ \hline
$N_{740}=317466$ & & & & &\\ \hline
$N_{840}=220250$ & & & & &\\ \hline
$N_{940}=131874$ & & & & &\\ \hline
$N_{050}=58170$ & $N_{051}=3544$ & $N_{052}=190$ & & &\\ \hline
$N_{150}=82790$ & $N_{151}=4156$ & $N_{152}=190$ & & &\\ \hline
$N_{250}=104070$ & $N_{251}=4380$ & & & &\\ \hline
$N_{350}=116486$ & $N_{351}=4156$ & & & &\\ \hline
$N_{450}=116486$ & $N_{451}=3544$ & & & &\\ \hline
$N_{550}=104070$ & & & & &\\ \hline
$N_{650}=82790$ & & & & &\\ \hline
$N_{750}=58170$ & & & & &\\ \hline
$N_{060}=21638$ & $N_{061}=1104$ & & & &\\ \hline
$N_{160}=26958$ & $N_{161}=1160$ & & & &\\ \hline
$N_{260}=30062$ & $N_{261}=1104$ & & & &\\ \hline
$N_{360}=30062$ & & & & &\\ \hline
$N_{460}=26958$ & & & & &\\ \hline
$N_{560}=21638$ & & & & &\\ \hline
$N_{070}=6888$ & $N_{071}=290$ & & & &\\ \hline
$N_{170}=7664$ & & & & &\\ \hline
$N_{270}=7664$ & & & & &\\ \hline
$N_{370}=6888$ & & & & &\\ \hline
$N_{080}=1916$ & & & & &\\ \hline
$N_{180}=1916$ & & & & &\\ \hline
       \end{tabular}
      \end{center}
     \end{table}



\clearpage
\normalsize
\section{Conclusion}

 We have shown in this paper that the free energy of topological sigma models
on $CP^{3}$,$CP^{4}$ and $Gr(2,4)$ (coupled to gravity but on small phase
 space) can actually be evaluated using DWVV equation, some properties of
topological field theory and one initial condition($\langle {\cal O}_{W_4}
{\cal O}_{W_4} \rangle =1$ for $CP^{3}$,$\langle {\cal O}_{W_5}
{\cal O}_{W_5} \rangle =1$ for $CP^{4}$ and $\langle {\cal O}_{W_5}
{\cal O}_{W_6} \rangle=1$ for $Gr(2,4)$). Although in $CP^{3}$ and $CP^{4}$
 case where classical cohomology ring of target space $M$ is generated by
$H^{2}(M)$, this fact was already shown by Kontsevich and Manin, we find
this strategy also work for $Gr(2,4)$. We also observe that for $CP^{3}$ and
$CP^{4}$, expansion
coefficients of the free energy acctually counts the number of rational curves
passing through $PD(W)$ of insertion operator ${\cal O}_{W}$ in $d=1$ case.
(see Appendix A) From (\ref{a9}) we see the contribution of one rational curve
satisfying ``passing through condition'' to correlation function is the
product of intersection numbers between  the rational curve and corresponding
Poincar\'e duals, however, it seems this factor equals  $1$, unless $W$ is the
K\"{a}hler form. If not so, most of correlation functions  in degree $d$
sector have to be devisible by $d$, but our results do not support this
speculation.
 In $Gr(2,4)$ case, we find a rather interesting symmetry in the correlation
functions. Interchange between the insertion of ${\cal O}_{W_3}$ and
${\cal O}_{W_4}$ (in words of Schubert cycle of $Gr(2,4)$,$W_{3}$ and $W_{4}$
 correspond to $\sigma_{1,1}$ and $\sigma_{2}$) does not change
correlation functions. This is natural because in classical geometry we can not
distinguish $\sigma_{1,1}$ from $\sigma_{2}$ in its algebraic structure.
In algebraic geometry, $Gr(2,4)$ describes lines in $CP^{3}$.
$\sigma_{2}$ correspond to the set of lines passing through a point $p_{0}$
and $\sigma_{1,1}$ to the set of lines contained fixed plane $h_{0}$. And
the symmeetry of the above two sets comes from duality, i.e we can see
$(a_1:a_2:a_3:a_4)$ as a point of $CP^{3}$ or as a linear form on it.
After the completion of this manuscript, a paper by Di Francesco and Itzykson
``Quantum intersection rings'' has appeared which overlaps with our work.

\vskip1cm
\noindent{\large\bf Acknowledgement}
\vskip0.5cm

 We would like to thank Prof.T.Eguchi for suggesting this problem and kind
encouragement.We also thank Dr.K.Hori for many useful discussions.
We are indebted to Dr.T.Hotta and Dr.T.Izubuchi for manipulation of computers.

\appendix
\section{Derivation of Initial Conditions and Some Direct
Counting of Amplitudes}
 We first show $\langle {\cal O}_{W_4} {\cal O}_{W_4}\rangle=1$ for $CP^{3}$
(resp. $\langle {\cal O}_{W_5} {\cal O}_{W_5} \rangle =1$ for $CP^{4}$).
{}From (\ref{a9}) this is just number of lines passing through two points
of $CP^{3}$(resp. $CP^{4}$), so it  equals to 1 trivially. But we derive
this result using schubert calculus of $Gr(2,4)$ (resp.$Gr(2,5)$) which
corresponds to the space of lines in $CP^{3}$ (resp.$CP^{4}$).
 Schubert cycles $\sigma_{a_1,a_2} \subseteq Gr(2,N)$ $(N-2 \geq a_1 \geq a_2
\geq 0)$ form a basis of $H^{*}(Gr(2,N),Z)$ and are given by following
definition.
\begin{equation}
 \sigma_{a_1,a_2} = \{l \in Gr(2,N)| dim_{C}(l \cap V_{N-2+i-a_i}) \geq i \}
\label{c1}
\end{equation}
 where $V_{i}$'s are linear subspace of $C^{N}$ of dimension $i$ satisfying
 following condition.
\begin{equation}
   V_1 \subset V_2  \subset \cdots \subset V_{N-1} \subset C^{N}
\label{c2}
\end{equation}
 From this definition, subset of $Gr(2,N)$ passing through a point of
$CP^{N-1}$ is given as $\sigma_{N-2,0}$ because this condition is equivalent
to $dim_{C}(l \cap V_{1}) = 1$. Then we can calculate $\langle {\cal O}_{W_4}
{\cal O}_{W_4} \rangle$ for $CP^{3}$ (resp.$\langle {\cal O}_{W_5}
{\cal O}_{W_5} \rangle$ for $CP^{4}$) as follows.
\begin{eqnarray}
 \langle {\cal O}_{W_4} {\cal O}_{W_4} \rangle = {}^{\sharp}(\sigma_{2,0}
\cdot \sigma_{2,0})_{Gr(2,4)} = {}^{\sharp}(\sigma_{2,2})_{Gr(2,4)} =1
\nonumber\\
 \langle {\cal O}_{W_5} {\cal O}_{W_5} \rangle = {}^{\sharp}(\sigma_{3,0}
\cdot \sigma_{3,0})_{Gr(2,5)} = {}^{\sharp}(\sigma_{3,3})_{Gr(2,5)} =1
\label{c3}
\end{eqnarray}
In this derivation, we used Pieri's formula
\begin{equation}
\sigma_{a,0} \cdot \sigma_{b_1,b_2} = \sum_{\stackrel{b_i \leq c_i \leq
b_{i-1}}{c_1+c_2=a+b_1+b_2}} \sigma_{c_1,c_2}
\label{c4}
\end{equation}
and $\sigma_{N-2,N-2}$ corresponds to a point of $Gr(2,N)$.

Next we derive $\langle {\cal O}_{W_5} {\cal O}_{W_6} \rangle =1$ for
$Gr(2,4)$. Using Pl\"ucker map, Gr(2,4) can be embedded in $CP^{5}$ as a
quadratic hypersurface. This embedding is constructed as follows.
 We map a line $\{\mbox{\boldmath$v$}_1,\mbox{\boldmath$v$}_2 \}_{C}$ in
$CP^{3}$($C^{4}$) to a multivector $\mbox{\boldmath$v$}_1 \wedge
\mbox{\boldmath$v$}_2 \in {\bigwedge}^{2}C^{4}$ This map (we call it
$\iota$) is injective and conversely the image of a line in ${\bigwedge}^{2}
C^{4}$ is characterized by decomposability, i.e. $\omega \in {\bigwedge}^{2}
C^{4}$ is in $Im(\iota)$ iff $\omega$ can be written as $\omega =
\mbox{\boldmath$v$}_1 \wedge \mbox{\boldmath$v$}_2$. It can be shown that
this condition is equiivalent to $\omega \wedge \omega =0$. So using a basis
$\{\mbox{\boldmath$e$}_1,\mbox{\boldmath$e$}_2,\mbox{\boldmath$e$}_3,
\mbox{\boldmath$e$}_4\}$ of $C^{4}$ and expanding $\omega$ as follows,
\begin{equation}
\omega = {\lambda}_{12}\mbox{\boldmath$e$}_1 \wedge \mbox{\boldmath$e$}_2
+ {\lambda}_{13}\mbox{\boldmath$e$}_1 \wedge \mbox{\boldmath$e$}_3
+{\lambda}_{14}\mbox{\boldmath$e$}_1 \wedge \mbox{\boldmath$e$}_4
+{\lambda}_{23}\mbox{\boldmath$e$}_2 \wedge \mbox{\boldmath$e$}_3
+{\lambda}_{24}\mbox{\boldmath$e$}_2 \wedge \mbox{\boldmath$e$}_4
+{\lambda}_{34}\mbox{\boldmath$e$}_3 \wedge \mbox{\boldmath$e$}_4
\label{c5}
\end{equation}
 we can realize $Gr(2,4) (\simeq Im(\iota))$ in $CP^{5}$ as follows.
\begin{eqnarray}
 &&\omega \wedge \omega = 0 \nonumber\\
&\Longleftrightarrow & {\lambda}_{12}{\lambda}_{34}-{\lambda}_{13}
{\lambda}_{24} +{\lambda}_{14}{\lambda}_{23} = 0
\label{c6}
\end{eqnarray}
 In summary, we can see $Gr(2,4)$ as a quadratic hypersurface $G$ in $CP^{5}$.
 Then we want to find the realization of $\sigma_{2,1}(=W_5)$ and
$\sigma_{2,2}(=W_6)$ in $G$. From the study of the structure of $G$
(see Chap.6 of Griffith Harris \cite{gh}) $\sigma_{2,1}$ corresponds to a line
in $G$ and trivially $\sigma_{2,2}$ to a point. If we consider plane $h$
(resp. line $l$) in $CP^{5}$, quadratic feature of $G$ makes the intersection
$(h \cap G)$ (resp.$(l \cap G)$) into conic of $G$
(resp. two points of $G$).Then we have to devide them by factor 2,i.e.
\begin{eqnarray}
 \sigma_{2,1} \leftrightarrow \frac 12 (h \cap G) \\
 \sigma_{2,2} \leftrightarrow \frac 12 (l \cap G)
\label{c7}
\end{eqnarray}
 The space of lines in $G$ (we denote it as $L_{G}$) is constructed as the
subspace of $Gr(2,6)$(space of lines in $CP^{5}$) using bundle calculation
(see \cite{jn}).
\begin{equation}
 L_{G} = c_{T}(Sym^{2}(U^{*}))=4 \tilde{\sigma}_{2,1}
\label{c8}
\end{equation}
 where $U$ is the universal bundle of $Gr(2,6)$.

(We denote schubert cycles of $Gr(2,6)$ as $\tilde{\sigma}_{a_1,a_2}$
in order to distinguish them from the ones of $Gr(2,4)$).

{}From (\ref{c7}), in $L_{G}$, to count the number of lines which passes
through $\sigma_{2,1}$ and $\sigma_{2,2}$ are equivalent to picking up
the lines which passes through $h$ and $l$
(multiplied by factor $\frac 12$).
Then we have
\begin{equation}
\langle {\cal O}_{W_5} {\cal O}_{W_6} \rangle =
{}^{\sharp}(\frac 12 \tilde{\sigma}_3 \cdot \frac 12 \tilde{\sigma}_{2}
\cdot 4 \tilde{\sigma}_{2,1}) =1
\label{c9}
\end{equation}
Lastly, using this technique, we calculate the topological amplitude of
$d=1$ sector for $CP^{3}$ and $CP^{4}$.
\begin{eqnarray}
\lefteqn{\underline{CP^3}}\nonumber\\
&&{\cal O}_{W_3}\leftrightarrow \sigma_1 \quad {\cal O}_{W_4}
\leftrightarrow \sigma_2 \quad (\mbox{in} Gr(2,4))\nonumber\\
&&\langle {\cal O}_{W_4}{\cal O}_{W_4} \rangle =
{}^{\sharp}(\sigma_2 \cdot \sigma_2) = 1 \nonumber\\
&&\langle {\cal O}^{2}_{W_3} {\cal O}_{W_4}\rangle =
{}^{\sharp}(\sigma^{2}_{1} \cdot \sigma_2) =1 \nonumber \\
&&\langle {\cal O}^{4}_{W_3}={}^{\sharp}(\sigma^{4}_{1})=2 \nonumber\\
\underline{CP^4}\nonumber\\
&&{\cal O}_{W_3} \leftrightarrow \sigma_1 \quad {\cal O}_{W_4} \leftrightarrow
\sigma_2 \quad {\cal O}_{W_5} \leftrightarrow \sigma_3
\quad (\mbox{in} Gr(2,5)) \nonumber\\
&&\langle {\cal O}_{W_5}{\cal O}_{W_5}\rangle =
{}^{\sharp}(\sigma_{3} \cdot \sigma_{3})=1 \nonumber\\
&&\langle {\cal O}^{3}_{W_3}{\cal O}_{W_5}\rangle =
{}^{\sharp}(\sigma^{3}_{1} \cdot \sigma_{3})=1 \nonumber\\
&&\langle {\cal O}^{3}_{W_4}\rangle =
{}^{\sharp}(\sigma^{3}_{2})=1 \nonumber\\
&&\langle {\cal O}_{W_3} {\cal O}_{W_4} {\cal O}_{W_5}\rangle =
{}^{\sharp}(\sigma_{1} \cdot \sigma_{2} \cdot \sigma_{3})=1 \nonumber\\
&&\langle {\cal O}^{2}_{W_3}{\cal O}^{2}_{W_4}\rangle =
{}^{\sharp}(\sigma^{2}_{1} \cdot \sigma^{2}_{2})=2 \nonumber\\
&& \langle {\cal O}^{4}_{W_3}{\cal O}_{W_4}\rangle =
{}^{\sharp}(\sigma^{4}_{1} \cdot \sigma_{2})=3 \nonumber\\
&&\langle {\cal O}^{5}_{W_3}\rangle =
{}^{\sharp}(\sigma^{5}_{1})=5
\end{eqnarray}


\begin{thebibliography}{99}
\bibitem{km}M.Kontsevich , Y.Manin.
\newblock{\em Gromov--Witten Classes,Quantum Cohomology,and Enumerative
 Geometry}
\newblock Commun.Math.Phys.164 (1994) 525;
\bibitem{dw}R.Dijkgraaf, E.Witten.
\newblock Nuclear Physics B342 (1990) 486;
\bibitem{itz}C.Itzykson.
\newblock{\em Counting rational curves on rational surfaces}
\newblock Saclay preprint \quad T94/001;
\bibitem{cv}S.Cecotti,C.Vafa.
\newblock Nuclear Physics B367 (1991) 359;
\bibitem{dub}B.Dubrovin.
\newblock{\em Geometry of 2D Topological Field Theory}
\newblock Preprint SISSA-89/94/FM \quad hepth \quad 9407018;
\bibitem{va}C.Vafa.
\newblock{\em Topological Mirrors and Quantum Rings}
\newblock HUTP-91/A059;
\bibitem{w1}E.Witten.
\newblock{\em Topological Sigma Models}
\newblock Commun.Math.Phys.118(1988)411;
\bibitem{w2}E.Witten.
\newblock{\em On the Structure of the Topological Phase of Two Dimensional
Gravity}
\newblock Nucl.Phys.B342(1990)486;
\bibitem{ey}T.Eguchi,S.K.Yang.
\newblock{\em $N=2$ Super Conformal Models as Topological Field Theories}
\newblock Mod.Phys.Lett.A Vol.5 No.21(1990)1693;
\bibitem{lcw}W.Lerche,C.Vafa,N.Warner.
\newblock{\em Chiral Rings in $N=2$ Super Conformal Field Theories}
\newblock Nucl.Phys.B324(1989)427;
\bibitem{h}K.Hori.
\newblock{\em Constraints For Topological Strings In $D \geq 1$}
\newblock UT-694;
\bibitem{jn}M.Jinzenji,M.Nagura.
\newblock{\em Mirror Symmetry and An Exact Calculation of $N-2$ Point
Correlation Function on Calabi-Yau Manifold Embedded in $CP^{N-1}$}
\newblock UT-6**;
\bibitem{gh}P.Griffith,J.Harris.
\newblock{\em Principles of Algebraic Geometry}
\newblock J.Wiley,N.Y.,1978;
\end{thebibliography}
\end{document}